\documentclass[aps,prc,twocolumn,superscriptaddress,showpacs]{revtex4}

\usepackage{graphicx}
\usepackage{latexsym}
\usepackage{bm}
\usepackage{amsmath}

\usepackage{color}

\newcommand{\comment}[1]{}

\newcommand{\gsim}{\mbox{\raisebox{-0.6ex}{$\stackrel{>}{\sim}$}}\:}
\newcommand{\lsim}{\mbox{\raisebox{-0.6ex}{$\stackrel{<}{\sim}$}}\:}

\begin{document}

\title{
Hydrodynamic afterburner for the Color Glass Condensate and
the parton energy loss
}

\author{Tetsufumi Hirano}
\affiliation{RIKEN BNL Research Center,
    Brookhaven National Laboratory, Upton, New York 11973, USA}
\author{Yasushi Nara}
\affiliation{%
Department of Physics, University of Arizona, Tucson, Arizona 85721, USA
}

\date{\today}
 
\begin{abstract}
We take hydrodynamic initial conditions
in relativistic heavy ion collisions
from the Color Glass Condensate (CGC)
picture
through the $k_T$ factorization formula.
Gluon distributions
produced from the CGC
are found to provide good initial conditions
for the hydrodynamic simulations
in Au + Au collisions at Relativistic Heavy Ion Collider
(RHIC) energies.
We reproduce the centrality, rapidity, and energy
dependences of multiplicity within this approach.
We also investigate the energy loss of
high $p_T$ partons in the dense thermalized medium
created from colliding two CGC's.
We find that
our results on
 the centrality dependence of nuclear modification factors
for pions and back-to-back correlation for charged hadrons
at midrapidity 
are consistent with the RHIC data up to semicentral events.
Whereas, our approach in which jets are calculated
from perturbative QCD $2\to2$ processes
predicts less suppression at forward rapidity region
compared to the BRAHMS data in Au+Au collisions at RHIC.
\end{abstract}

%\pacs{24.85.+p,25.75.-q, 24.10.Nz}

\maketitle

\section{Introduction} 

Experiments of
high energy $pp$, $dA$, and $AA$ collisions are currently performed
at the Relativistic Heavy Ion Collider (RHIC) in Brookhaven
National Laboratory
for comprehensive understanding of the highly hot/dense matter,
the quark gluon plasma (QGP).
At collider energies, collisions of two relativistic
nuclei involve many aspects of physics
according to the relevant energy or time scale.
There already exist many theoretical approaches to understand
numerous RHIC data.
We consider in this work, particularly, 
the physics of the gluon saturation in a colliding nucleus,
hydrodynamic evolution of produced matter,
and the energy loss of hard partons in the medium.
Our goals are to combine them and to take a first step
toward a unified
understanding of the dynamical aspect of
high energy heavy ion collisions.

First data reported by the STAR Collaboration at 
RHIC~\cite{STAR:v2_130} revealed that
the observed large magnitude of elliptic flow
for hadrons
is naturally explained by hydrodynamics.
This suggests that
large pressure possibly in the partonic phase
is built at the very early stage
($\tau \sim 0.6$ fm/$c$)
in Au+Au collisions at RHIC.
This is one of the strongest indications of early
thermalization
of the QGP at RHIC~\cite{Huovinen:2003fa,TH:qm2004}.
These hydrodynamic predictions also give good agreements
with mass dependences of the second harmonic coefficient of
azimuthal distribution $v_2(p_T)$ in the low $p_T$ region~\cite{KHHH}.
This is in contrast to the case in the lower energy
collisions where the hydrodynamics always overpredicts
the data~\cite{NA49:v12,Agakichiev:2003gg}.
Hydrodynamics also predicts that the scaled elliptic flow,
which is defined as the second harmonics $v_2$ divided by
initial spatial eccentricity $\varepsilon$,
becomes almost constant around 0.2~\cite{Kolb:2000sd}.
The experimental data reaches the hydrodynamic limit
for the first time
in central and semicentral collisions at RHIC energies \cite{Adler:2002pu}.
Moreover, $v_2$ as a function of pseudorapidity~\cite{PHOBOS:v2}
can be reproduced by hydrodynamics
only in $\mid \eta \mid < 1$~\cite{Hirano:2001eu}.
All these analyses indicate that
a high dense matter created in heavy ion collisions at RHIC energies
achieves (local) thermal equilibrium state
in the low $p_T$ region, around midrapidity, and up to semicentral
collisions.

On the other hand,
jet production rate becomes large enough to measure
high $p_T$ power-law tails at RHIC.
These minijets
go across the expanding matter
rather than participate
 in the dense medium
in the case of heavy ion collisions.
During traveling through the expanding medium,
energetic partons interact with
soft matter and lose their energies dynamically
(jet quenching).
Thus, high $p_T$ hadrons coming from
fragmentation of minijets can be good probes
of the bulk matter~\cite{GVWZKW}.
It was observed that
nuclear modification factors for hadrons % do not say charged hadrons
in the high $p_T > 5$ GeV/$c$ region are
considerably smaller than unity in central
collisions~\cite{phenix:pi0,star:highpt,Back:2003qr,BRAHMS:dA}
and that it increases with decreasing
centrality and eventually reaches around 
unity in peripheral collisions.
Disappearance of the away-side peak in azimuthal
correlation functions for high $p_T$ hadrons is also
observed in central Au+Au collisions~\cite{STAR:btob}.
This can be described by perturbative QCD (pQCD)
based parton models with parton energy loss
in the dense medium~\cite{Wang:2003mm,Wang:2003aw,HN2}.
Cronin enhancement of the hadron spectra and 
existence of the back-to-back
correlation at mid-rapidity in $dA$ collisions
at RHIC~\cite{PHOBOS:dA,PHENIX:dA,STAR:dA,BRAHMS:dA}
support that
suppression of the yield
and disappearance of the away-side peak
in Au+Au collisions at mid-rapidity
are due to the final state interaction.
The radiative parton energy loss induced by the medium has been
studied extensively
~\cite{BDMPS,Zakharov,Wiedemann,GLV,WW,Arnold:2002ja}
and found that
the energy loss is proportional to the gluon rapidity density of the medium.
Theoretical analyses reveal that
a large parton density is required 
to account for observed suppressions of single as well as
dihadron spectra at RHIC~\cite{Wang:2003aw,VG,Salgado:2003gb}.

From the above discussion,
the current RHIC data strongly
suggest that the parton density
created in heavy ion collisions at RHIC
is dense enough to cause both large elliptic flow
of bulk matter
and large suppression of high-$p_T$ hadrons.
One of the important problems in the physics of
heavy ion collisions is to reveal the origin 
of this dense matter.
The bulk particle production in high energy hadronic collisions
 is dominated by the small $x$
modes in the nuclear wave function, where $x$ is a momentum
fraction of the incident particles.
Therefore, an origin of the large density could be traced to the initial
parton density at small $x$ inside the ultra-relativistic nuclei
 before collisions~\cite{AlMueller1999}.
It is well known that gluon density increases rapidly 
with decreasing $x$ by the BFKL cascade
until gluons begin to overlap in phase space~\cite{GLR83}
where nonlinear interaction becomes important.
These gluons eventually form the Color Glass Condensate (CGC)~\cite{Iancu:2003xm}.
This phenomenon is characterized by a ``saturation scale"
$Q_s^2$ which corresponds to the gluon rapidity
density per unit transverse area.
When $Q_s^2$ is large ($Q_s\gg\Lambda_{\mathrm{QCD}}$),
the coupling becomes weak ($\alpha_s(Q_s)\ll 1$).
In addition, the gluon occupation number in the wave function
is huge, $\sim 1/\alpha_s(Q_s) \gg 1$.
Therefore this can be studied by an weak coupling classical method
known as McLerran-Venugopalan (MV) model~\cite{MV}.
It has been developed to compute the classical parton distributions
of nuclei based on this model.
Moreover, renormalization group methods
which systematically
incorporate quantum corrections to
the classical effective theory were developed~\cite{RGE1,RGE2,RGE3}.
There is an evidence for gluon saturation 
in deep inelastic scattering at HERA.
Models based on the saturation picture fit the HERA data well
in the moderate $Q^2$ and low $x$~\cite{GBW,IIMu03}.
This effective theory can be applied to the collisions of nuclei
and its solution can be used for initial conditions of
further evolution of the system created
in heavy ion collisions~\cite{MBVDG,BMSS,BMSS2,Shin}.
Perturbative solutions of
the gluon production for collisions of two nuclei~\cite{KMW,KR}
as well as $pA$ collisions~\cite{Kovchegov:1998bi,Dumitru:pA}
 have been computed.
Non-perturbative classical solutions of the Yang-Mills equations
were also obtained by the real-time numerical simulations
on the lattice~\cite{KV,KNV,Lappi}.
The calculations based on the CGC picture have been compared to
 various RHIC data.
Remarkably, the CGC results on the global observables, e.g.,
the centrality, rapidity and energy dependences of charged
hadron multiplicities agree with the RHIC data under the assumption
of parton-hadron duality~\cite{KLN}.
The CGC picture is also applied for studies of particle correlations
 and azimuthal
anisotropies~\cite{KT}, and valence quark distributions at small-$x$
~\cite{Itakura:2003jp}.
% It is also applied to study particle correlations and azimuthal
% anisotropies~\cite{KT}, and valence quark distributions at small-$x$
% ~\cite{Itakura:2003jp}.
It has been shown that
the classical wave function in the MV model contains
Cronin enhancement
and that quantum evolution with respect to $x$ makes
 the spectrum suppressed~\cite{BKW,croninCGC}.
Suppression of nuclear modification factor
 at forward rapidity region in $dA$ collisions~\cite{Jalilian-Marian:2004xm}
in the CGC picture
is also consistent with
the recent RHIC data~\cite{BRAHMS:qm2004}.

The above facts show that
the CGC, hydrodynamics, and the energy loss of hard partons
 are the key ingredients to describe the RHIC physics.
One expects that they are indeed
closely related with each other.
For example, the CGC could be a good initial condition
for a thermalized state
because it produces a large number of gluons.
Thus a large number of gluons are responsible for large suppression
of jet spectra.
In order to understand systematically
the whole of the stages
of relativistic heavy ion collisions,
it is indispensable to construct a unified and dynamical
framework in which the above different ideas
are consistently incorporated.
In this work, we perform hydrodynamic simulations with the initial
condition taken from the CGC and simulate parton energy loss
in the fluid elements.
This approach will lead us to get deeper understanding of the
dynamical aspect of heavy ion collisions.

In addition, some of the problems which are inherent
in a particular
approach can be removed.
For instance, 
one conventionally
parametrizes initial conditions in the hydrodynamic simulations
shortly after a collision of two nuclei so as to
reproduce final observed spectra.
Hydrodynamic framework itself does not provide an initial
condition.
Hence one has to choose it 
from infinite possible candidates.
Any initial conditions
which reproduce the final hadron spectra are equivalent,
so it is hard to discard even one of them
within the hydrodynamic approach.
This is the ambiguity problem of hydrodynamic model
in heavy ion collisions.
A good example
can be found in Ref.~\cite{Huovinen:1998tq}:
Two completely different initial conditions end up
almost the same rapidity distribution.
Therefore it is desired to take an initial condition which
is obtained by a reliable theory.
On the other hand, most of the calculations based on
the CGC do not 
include final state interactions.
According to estimation of initial parton productions
 from a classical lattice Yang-Mills
simulation based on the CGC~\cite{Lappi,KNV3},
the transverse energy per particle
is roughly $E_T/N \sim 1.5$ GeV,
while the experimental data yields $E_T/N \sim 0.6$ GeV \cite{PHENIX:meanet}.
The elliptic flow from a classical Yang-Mills simulation is
inconsistent with RHIC data~\cite{KNV2}.
Those discrepancies, which are mainly due to the lack
of collective effects, will be removed by introducing further time
evolutions through, e.g., hydrodynamics.

For the calculation of parton energy loss, 
one needs time evolution of parton density.
Bjorken expansion \cite{Bjorken:1982qr} is assumed 
for the evolution of parton density
in almost all work except Ref.~\cite{Gyulassy:2001kr}
in which hydrodynamic simulations are used
to estimate the energy loss of partons for the first time.
We have previously established a model (the hydro+jet model)
in which hydrodynamic evolutions are combined with
explicitly traversing 
non-thermalized partons
with large transverse momenta~\cite{HN1,HN2,HN3,HN4}.
Here the production of high $p_T$ partons
is described by a pQCD parton model.
These high $p_T$ partons come from parton distribution function
in relatively large $x$ region.
Systematic studies based on the hydro+jet model have been performed
for $p_T$ spectra \cite{HN4},
back-to-back correlation functions~\cite{HN2},
pseudorapidity dependence of nuclear modification factors~\cite{HN3},
and elliptic flow \cite{HN3,HN4}.
For the initial condition of hydrodynamic simulations,
we simply parametrized the initial shape of energy density
in the previous work.
Instead, we will employ the CGC picture 
and use gluons produced from melting CGC
for initial conditions for hydrodynamic simulations
in this paper.
The number and the energy density distributions
for produced gluons
are calculated from the $k_T$ factorization formula~\cite{KLN}.
Although the produced gluons
will reach a thermalized state through the dissipative
processes in the realistic situations,
the description of non-equilibrium phenomena
is beyond the scope of the present paper.
Instead, we assume thermalization for gluons produced from the CGC.

This paper is organized as follows. In Sec.~\ref{section:model},
we discuss how the gluon distribution produced from the CGC
is used for the initial condition in the hydrodynamics.
In Sec.~\ref{sec:hydrojet}, we briefly summarize the hydro+jet model.
In Sec.~\ref{sec:result}, we present hydrodynamic results
on (pseudo)rapidity distributions for charged hadrons
 and the transverse momentum distributions for pions, kaons, and protons.
We  show that
the CGC provides good initial conditions
in the hydrodynamic model.
We  also show that,
with almost keeping the shape of rapidity distribution
during the hydrodynamical evolution,
transverse collective flow is generated
by pressure gradient perpendicular to the collision axis.
We study the centrality dependence of the nuclear modification
factor for neutral pions and the back-to-back correlations
for charged hadrons.
The previous analysis on the high $p_T$ hadron spectra in the forward
rapidity region is revisited.
In the final section, we give a summary and a discussion about
the further improvements in our approach.

\section{Initial Condition for hydrodynamics from gluon saturation}
\label{section:model}

There exists a lot of extensive work for the description of gluon productions
in nuclear collisions
in the saturation regime where nonlinear effects
become important. 
Perturbative solutions for the collision
of two nuclei in the MV model were obtained in Refs.~\cite{KMW,KR}.
However, those solutions are limited for the relatively high momentum
and cannot be used for the calculation of total gluon multiplicities.
In Ref.~\cite{Kovchegov:2000}, analytic expression for the
gluon multiplicity is obtained.
Non-perturbative classical boost invariant solutions
are currently
available on the lattice \cite{KV,KNV,KNV3,Lappi}
and it is highly desired to use the lattice results
for obtaining reliable initial conditions.
However, one of our purposes in this paper
is to investigate the rapidity dependence
of final hadrons.
Therefore,
we employ the $k_T$ factorization formula along
the line of work done by Kharzeev, Levin, and Nardi (KLN)~\cite{KLN}
for the computation of the gluon rapidity distribution
which is to be the initial condition
for sequential hydrodynamic evolution.

\subsection{Gluon Production in the $k_T$ factorization formula}

The number of produced gluons
in the $k_T$-factorization formula
 is given by~\cite{GLR81,GLR83,LL94,Szczurek:2003fu}
\begin{eqnarray}
\frac{dN_g}{d^2x_{\perp}dy}&=&
   \frac{4\pi^2N_c}{N_c^2-1} \int\frac{d^2p_T}{p^2_T}
   \int d^2k_T \alpha_s(Q^2)   \nonumber\\
 &\times&      \phi_A(x_1,k_T^2;\bm{x}_\perp)
               \phi_B(x_2,(p_T-k_T)^2;\bm{x}_\perp), \nonumber \\
\label{eq:ktfac}
\end{eqnarray}
where
$x_{1,2}=p_T\exp(\pm y)/\sqrt{s}$ and $y$ and
$p_T$ are a rapidity
and a transverse momentum of a produced gluon.
Running coupling $\alpha_s$ is evaluated at the scale
$Q^2 = \max(k^2_T,(p_T-k_T)^2)$.
The unintegrated gluon distribution $\phi$ is related to
the gluon density of a nucleus by
\begin{equation}
   xG_A(x,Q^2) = \int^{Q^2} d^2k_T d^2x_\perp \phi_A(x,k^2_T;\bm{x}_\perp).
\end{equation}
There are several models/parametrizations
in the literature which
provide a gluon distribution in the saturation region.
In principle,
the $x$ dependence of 
unintegrated distribution functions
in Eq.~(\ref{eq:ktfac})
should be solutions to the 
nonlinear quantum evolution
equations such as %BFKL (Balitsky-Fadin-Kuraev-Lipatov) equation or
Balitsky-Kovchegov equation~\cite{Balitsky:1995ub,Kovchegov:1999yj}
or, more generally, JIMWLK equation~\cite{RGE1,RGE2,RGE3}.
Leaving this analysis for the future work,
we follow basically the KLN approach
which captures the essential features of
the gluon saturation physics
in a simple manner.

In the MV model~\cite{MV},
the gluon distribution below the saturation scale
$Q_s^2$
is logarithmically suppressed
$\phi(k_T^2) \sim \ln(Q_s^2/k_T^2)$
compared to the perturbative distribution $\sim 1/k_T^2$.
Instead, motivated by the KLN approach, we use a simplified
assumption about the unintegrated gluon distribution
function:
\begin{equation}
\label{eq:uninteg}
\phi_A(x,k^2_T;\bm{x}_\perp)\,\,
  =\left\{\begin{array}{l}
   \frac{\kappa C_F}{2\pi^3\alpha_s(Q^2_s)}\frac{Q_s^2}{Q_s^2+\Lambda^2}, \,                      \quad  k_T\,\leq\,Q_s, \\
   \frac{\kappa C_F}{2\pi^3\alpha_s(Q^2_s)}\, \frac{Q^2_s}{k^2_T+\Lambda^2},
              \quad k_T\,>\,Q_s,
\end{array}
\right.
\end{equation}
where $C_F=(N_c^2-1)/(2N_c)$. We introduce a small regulator
$\Lambda=0.2$ GeV/$c$ in order to
have a smooth distribution in the forward rapidity region
$|y|>4.5$ at RHIC.
Other regions are not affected by introducing a small regulator.
As we will discuss below, 
the above distribution depends on the transverse
coordinate $\bm{x}_\perp$ through $Q_s^2$.
We are interested in the centrality and
rapidity dependences of 
total yields which are dominated by
the low transverse momenta.
The above simple form
should be fine for this purpose,
although 
the effect of anomalous dimension resulting
from the presence of the CGC
modifies
the $p_T$ dependence of the produced
gluons
as shown in Ref.~\cite{IIM02,KLM03,KLM04,IIM03,BKW}.
An overall constant factor $\kappa$ is determined by fitting
the multiplicity of charged hadron 
at midrapidity at $\sqrt{s_{NN}}=200$ GeV for the most central collision.
Saturation momentum of a nucleus $A$
in $A+B$ collisions is obtained by solving the following
implicit equation with respect to $Q_s$
at fixed $x$ and $\bm{x}_{\perp}$
\begin{equation}
\label{eq:saturation}
 Q^2_s(x, \bm{x}_{\perp}) = \frac{2\pi^2}{C_F}
\alpha_s(Q^2_s)xG(x,Q^2_s)
              \rho^A_{\mathrm{part}}(\bm{x}_{\perp}),
\end{equation}
where 
\begin{equation}
   \rho^A_{\mathrm{part}}(\bm{x}_{\perp}) = T_A(\bm{x}_{\perp})
 \left\{1 - [1-
\sigma^{\mathrm{in}}_{NN}T_B(\bm{x}_{\perp})/B ]^B
          \right\}.
\end{equation}
The number of participants is given by
\begin{equation}
   N_{\mathrm{part}} = \int d^2x_{\perp}
           ( \rho^A_{\mathrm{part}}(\bm{x}_{\perp})
           +\rho^B_{\mathrm{part}}(\bm{x}_{\perp}) ).
\end{equation}
We take a simple perturbative form
for the gluon distribution for a nucleon
\begin{equation}
\label{eq:xG}
  xG(x,Q^2) = K\ln\left( \frac{Q^2 + \Lambda^2}{\Lambda_{\mathrm{QCD}}^2}\right)x^{-\lambda} (1-x)^n
\end{equation}
where
$\Lambda = \Lambda_{\mathrm{QCD}}=0.2$ GeV.
$K$ is used to control
saturation scale in Eq.~(\ref{eq:saturation})~\cite{BKW}.
We choose $K=0.7$  for $\lambda=0.2$
($K=0.55$ for $\lambda=0.25$ or $K=0.44$ for $\lambda=0.3$)
so that
average saturation momentum in the transverse plane yields
$\langle Q_s^2(x=0.01)\rangle \sim 2.0$ GeV$^2/c^2$
in Au+Au collisions at impact parameter $b=0$ at RHIC.
Similar to the KLN approach, $x^{-\lambda}$ dependence of
the saturation scale
is motivated
by the Golec-Biernat--W\"usthoff model~\cite{GBW}.
The factor $(1-x)^n$ shows
that gluon density becomes small at $x\to1$.
$n$ usually depends on the scale $Q^2$.
Here we take $n=4$ as in the KLN approach~\cite{KLN}.

We obtain the rapidity distribution for produced gluons
at each transverse point $\bm{x}_\perp$
by performing the integral of Eq.~(\ref{eq:ktfac})
numerically.
The transverse energy distribution
$dE_T/dy$ is also obtained
 by weighting the transverse momentum of gluons
$p_T$ in the integration with respect to $p_T$
in Eq.~(\ref{eq:ktfac}).
We cut off the integral range of $p_T$ 
in Eq.~(\ref{eq:ktfac}) since
only the low $p_T$ partons are assumed to
reach the local thermal equilibrium.
Instead, hard partons are
included by assuming usual pQCD $2\to2$ processes
in which DGLAP evolution is included as it is important in
the high $Q^2$ reactions.
It is unclear which value we have to use for
the cut off momentum $p_{T,\mathrm{cut}}$,
because it contains a nonperturbative nature
and also it would depend on the choice of unintegrated wave function.
We set the cut off momentum at $p_{T,\mathrm{cut}}=3$
GeV/$c$ which corresponds
roughly to the maximum saturation scale at $x=0.01$
at the origin $\bm{x}_{\perp} = \bm{0}$
in central Au+Au collisions at RHIC.
%%%%%%%%%%%%%%%%%%%%%%%%%%%%%%%%%%%%%%%%%%%%%%%%%%%%%%%%%%%%
% dndy
%%%%%%%%%%%%%%%%%%%%%%%%%%%%%%%%%%%%%%%%%%%%%%%%%%%%%%%%%%%%
\begin{figure}[ht]
\includegraphics[width=3.3in]{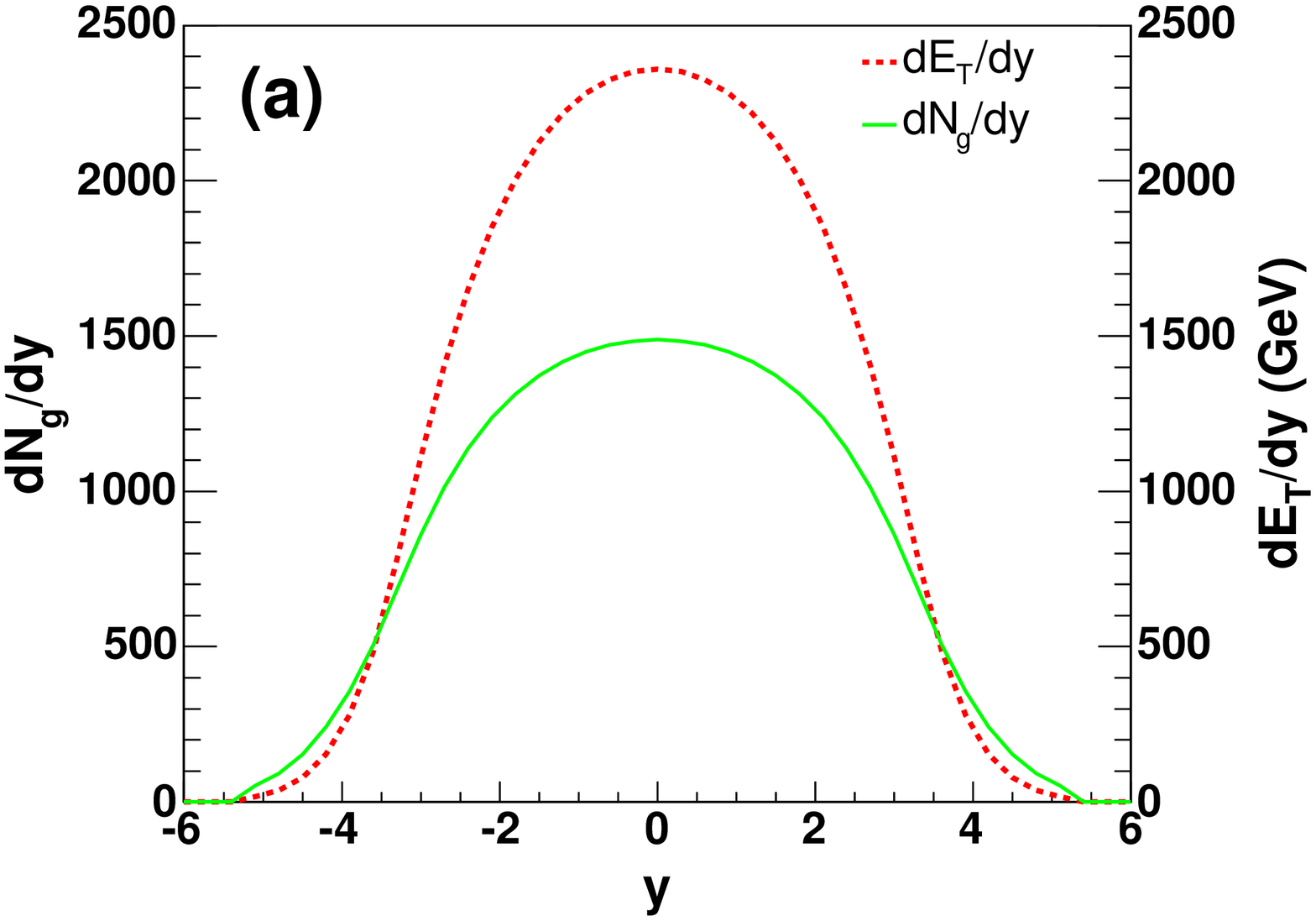}
\includegraphics[width=3.3in]{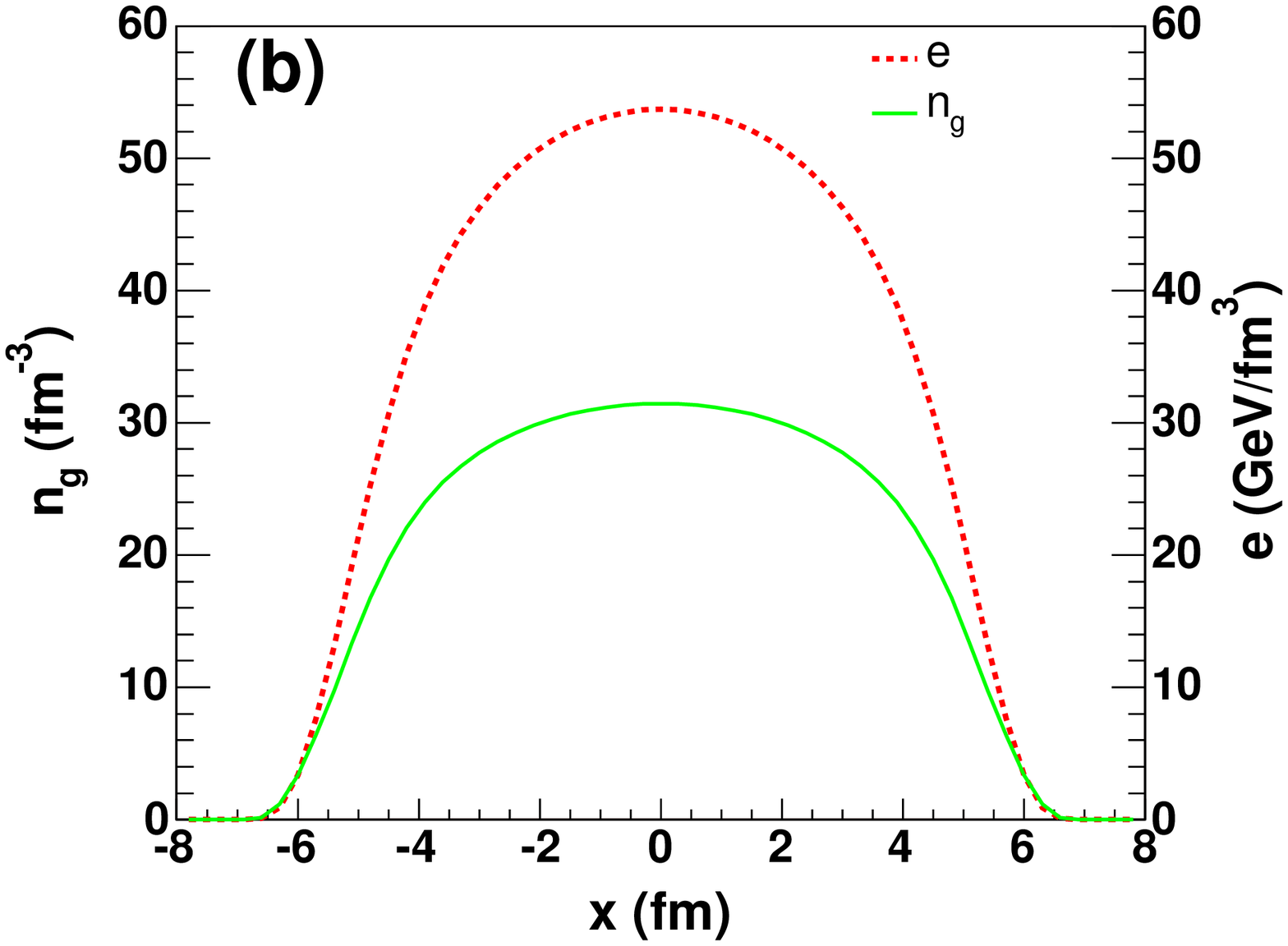}
\caption{
(a) Rapidity dependence of initial gluon transverse energy (dashed line)
and number distribution (solid line)
 in Au + Au  collisions at $\sqrt{s_{NN}}=200$ GeV
at $b=2.4$ fm.
(b) Gluon number and energy densities as functions of a transverse
coordinate.
Parameters are $\kappa^2=3.6$, $K=0.7$, and $\lambda = 0.2$.  
As we will discuss in the next subsection,
these parameters correspond to an initialization of
hydrodynamic simulations
which matches the number density of gluons 
produced by collisions of two CGC's
with the parton distribution in hydrodynamic simulations
at initial time.
The value of $\kappa^2$ is chosen so that we reproduce
the multiplicity of final hadrons as discussed
in Sec. IV.
In this procedure, the transverse energy distributions
represented by dashed lines are not used
as hydrodynamic initial conditions.
In the next subsection, we will
also discuss the other matching procedure
in which the transverse energy distribution is used
instead of the gluon number density.
In that case, the gluon distributions are 
factor 1.6 smaller than the results in these figures.
}
\label{fig:dndy_g}
\end{figure}
%%%%%%%%%%%%%%%%%%%%%%%%%%%%%%%%%%%%%%%%%%%%%%%%%%%%%%%%%%%%

Figure \ref{fig:dndy_g} (a)
shows rapidity distributions of produced gluons
obtained from Eq.~(\ref{eq:ktfac}).
Assuming free streaming in the longitudinal direction
until $\tau=0.6$ fm/$c$, we also obtain the number density
and the energy density as functions of
 a transverse coordinate
as shown in Fig.~\ref{fig:dndy_g} (b).
Here the direction of the $x$-axis is the same
as that of the impact parameter vector.
Solid (dashed) line corresponds to 
the number (energy) distribution
of produced gluons.

Regardless of the range of the parameter $\lambda=0.2$-0.3,
we find $k_T$ factorization formula gives almost the same shape
of rapidity distribution.
The initial transverse energy per particle is estimated to be
$E_T/N_g \sim 1.6$ GeV at $y=0$.
This is within a range estimated
in a numerical simulation of
the classical Yang-Mills equation~\cite{Lappi,KNV3}.

\subsection{How to connect the CGC to hydrodynamics}

In order to obtain a correct initial condition for hydrodynamics,
one needs non-equilibrium description
for the collisions of heavy nuclei.
%just after the collisions.
Especially, whether
 thermalization is achieved 
within a few fm/$c$ is
a longstanding problem in the physics of heavy ion collisions.
The Boltzmann equation with only the elastic parton-parton scattering
predicts that the system never reaches
a thermal state within a reasonable time scale
in heavy ion collisions~\cite{MBVDG}.
Inelastic processes such as
$gg \leftrightarrow ggg$
are quite responsible
for the equilibration of the system~\cite{BMSS,Shin}.
Although the thermalization process can alter
initial conditions,
the centrality dependence of total hadron multiplicities
is approximately preserved in the CGC initial conditions~\cite{BMSS2}.
QCD plasma instabilities may also 
play an important role in the thermalization
of matter from anisotropic initial conditions and
these may help to speed up the equilibration of the plasma~\cite{Randrup}.
A question of chemical equilibration has been addressed in Refs.
~\cite{Shuryak92,Biro:1993qt}.

%%%%%%%%%%%%%%%%%%%%%%%%%%%%%%%%%%%%%%%%%%%%%%%%%%%%%%%%%%%%
% de/dy
%%%%%%%%%%%%%%%%%%%%%%%%%%%%%%%%%%%%%%%%%%%%%%%%%%%%%%%%%%%%
\begin{figure*}[t]
\includegraphics[width=3.3in]{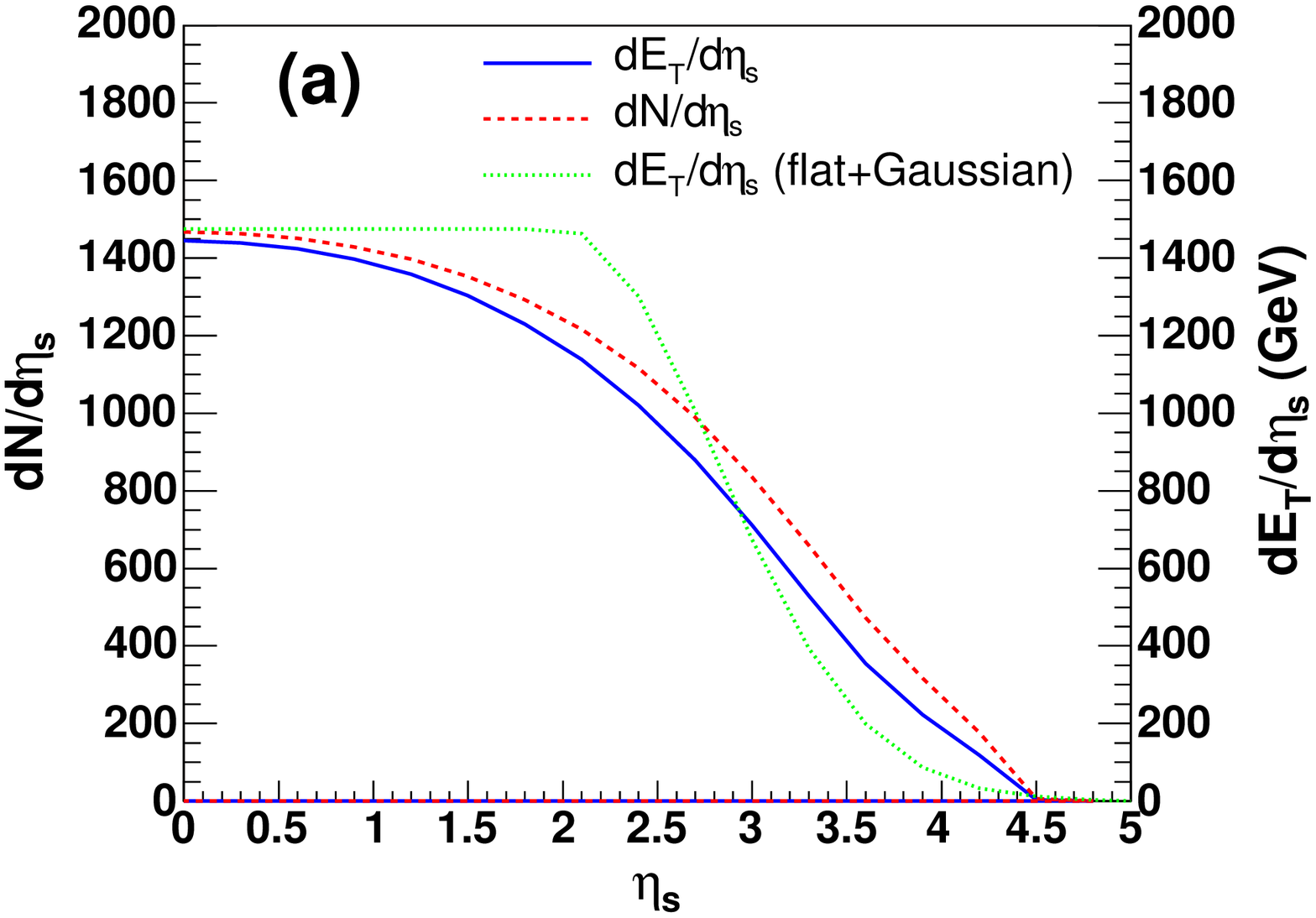}
\includegraphics[width=3.3in]{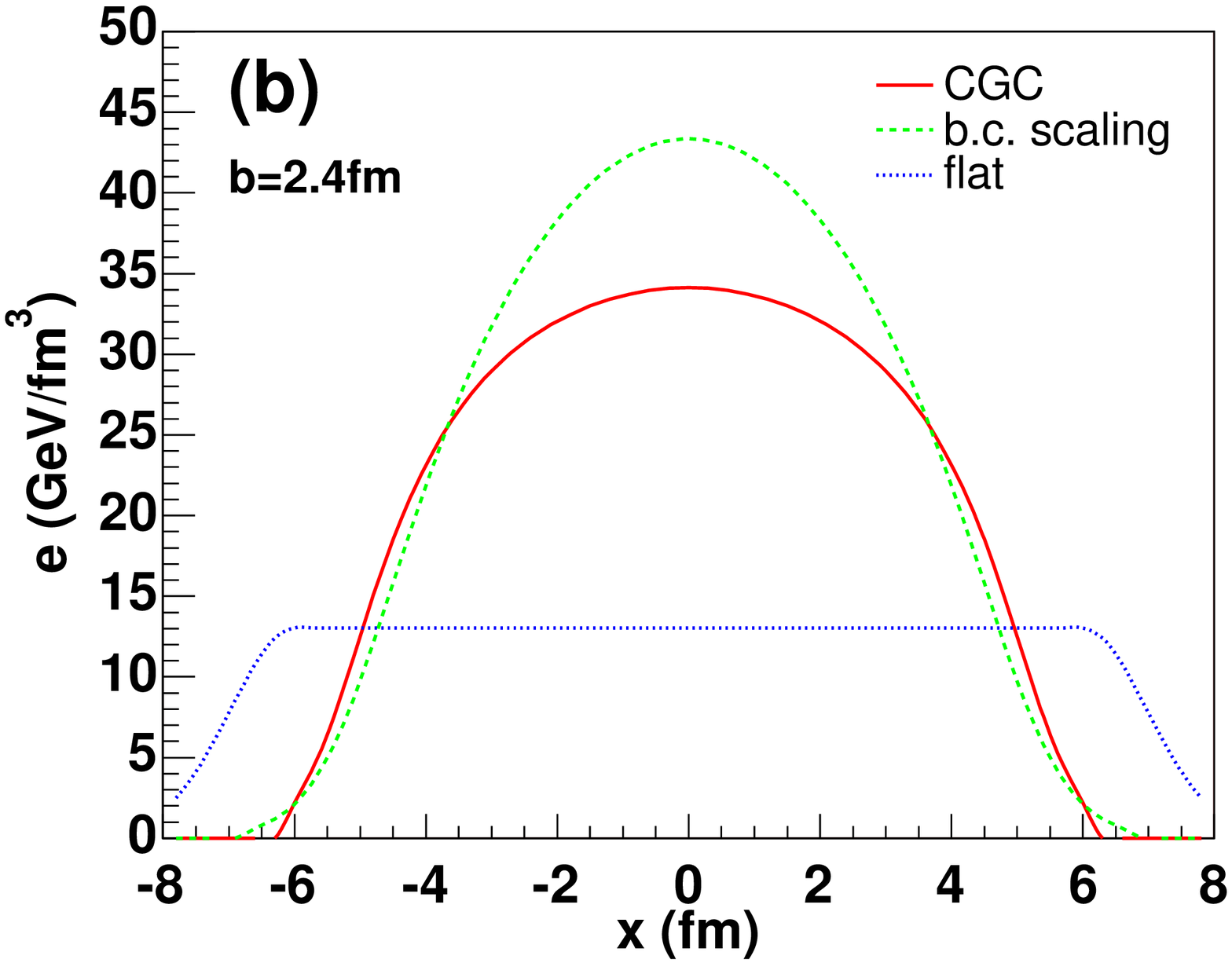}
\caption{
(a) Space-time rapidity dependence of the
initial conditions $dN/d\eta_s$ and $dE_T/d\eta_s$
in Au + Au  collisions at $\sqrt{s_{NN}}=200$ GeV.
%as obtained from the CGC after assuming thermalization.
We also compare the initial energy density distribution
with the one which we employed previously in Ref.~\cite{HiranoTsuda}.
(b) Comparison of the transverse profile from IC-$n$
with those in Refs.~\cite{HiranoTsuda} and \cite{Morita}.
}
\label{fig:ini_dens}
\end{figure*}
%%%%%%%%%%%%%%%%%%%%%%%%%%%%%%%%%%%%%%%%%%%%%%%%%%%%%%%%%%%%
Leaving these problems for the future work,
we assume that the system of gluons initially produced
from the CGC
reaches a kinematically as well as chemically
local equilibrate state at a short time scale.
We further assume that during being thermalized, 
the shape of the rapidity distribution is not changed.
Thus, we take initial conditions
from gluon distribution
obtained in the previous subsection
based on the CGC.

We consider two simple ways
to provide initial conditions from the CGC,
i.e., matching of number density  (IC-$n$)
and matching of energy density (IC-$e$).
We will compare charged particle distributions from these
initial conditions with each other in Sec.~\ref{sec:dndy}.
Assuming Bjorken's ansatz $y=\eta_{\mathrm{s}}$ \cite{Bjorken:1982qr}
where $\eta_{\mathrm{s}}$
is the space-time rapidity $\eta_{\mathrm{s}}=(1/2)\ln(x^+/x^-)$, 
we obtain the number density and the energy density
for gluons at a space-time point
$(\tau_0, \bm{x}_\perp, \eta_{\mathrm{s}}) \equiv(\tau_0, \vec{x})$
from Eq.~(\ref{eq:ktfac})
\begin{eqnarray}
\label{eq:g_density}
n_g(\tau_0,\vec{x}) = \frac{dN_g}{\tau_0 d\eta_{\mathrm{s}} d^2x_\perp},\\
\label{eq:et_density}
e_g(\tau_0,\vec{x}) = \frac{dE_T}{\tau_0 d\eta_{\mathrm{s}} d^2x_\perp}.
\end{eqnarray}
We take two steps
to specify initial conditions in hydrodynamic
simulations by using Eqs.~(\ref{eq:g_density})
or (\ref{eq:et_density}).
Firstly, we assume that these partons are thermally
equilibrated.
Secondly, the gluon number
density simply represents the 
$N_f = 3$ parton
number density, i.e., the QGP.
Although the initial time $\tau_0$ could depend on incident
energies (or even on the space coordinate through the saturation scale),
we assume a common $\tau_0$ for $\sqrt{s_{NN}}=130$ and 200 GeV.
We note that final particle multiplicity is not sensitive
to the initial time $\tau_0$ within a range
$ 0.5 \leq \tau_0 \leq 1.0$ fm/$c$
compared to the energy dependence
of the number of initial gluons.

Let us recall the relations among thermodynamical variables, i.e., 
energy density $e$, temperature $T$, and number density $n$
for the massless free parton system:
\begin{eqnarray}
  n &=& \left( \frac{3}{4}d_q + d_g \right) \frac{\zeta(3)}{\pi^2}T^3, \\
  e &=& \left( \frac{7}{8}d_q + d_g \right) \frac{\pi^2}{30}T^4 + B,
\end{eqnarray}
where $d_q=2N_cN_sN_f=36$, $d_g=2(N_c^2-1)=16$,
$\zeta(3)=1.20206$, and $B$(= 486 MeV/fm$^3$) is a bag constant
which is fixed by matching pressure between the QGP phase
and the hadron phase.
Thus,
we obtain the initial condition in the three-dimensional (3D) space
from the number density (IC-$n$):
\begin{eqnarray}
T(\tau_0, \vec{x}) = \left\{\frac{\pi^2 n(\tau_0, \vec{x})}
{43\zeta (3)} \right\}^{1/3},
\end{eqnarray}
or, from the energy density (IC-$e$)
\begin{equation}
T(\tau_0, \vec{x}) = \left\{  \frac{30[e(\tau_0, \vec{x})-B]}{47.5\pi^2}\right\}^{1/4}.
\end{equation}

When the temperature at ($\tau_0$, $\vec{x}$)
is below the critical temperature $T_{\mathrm{c}}=170$ MeV,
the all thermodynamic variables in the site are set to zero.
Figure~\ref{fig:ini_dens} (a) shows
the initial parton and energy
distributions
at the initial time $\tau_0=0.6$ fm/$c$
as a function of space-time rapidity $\eta_s$.
Here $dN/d\eta_{s} =\tau_0 \int d^2x_\perp n(\tau_0,\vec{x})$
and $dE_{T}/d\eta_{s} =\tau_0 \int d^2x_\perp e(\tau_0,\vec{x})$.
It should be noted that the assumption of the thermalization
for gluons produced from the CGC
is to reduce the transverse energy per particle from
$E_T/N_g=1.6$ GeV to $E_T/N_g \sim 1$ GeV
within our present parameters.
This change should be obtained by non-equilibrium descriptions
of pre-thermalization stages.
We also plot an initial condition
which was employed previously in Ref.~\cite{HiranoTsuda}.
Here we integrate the energy density
over all fluid elements with $T>100$ MeV.
The previous initial condition also
reproduces the rapidity distribution of final charged hadrons.
For longitudinal profiles, it consisted
of two regions, i.e., 
boost invariant region near midrapidity $|\eta_{\mathrm{s}}'|<2$
and Gaussian shape in forward/backward rapidity regions.
Here $\eta_{\mathrm{s}}'$ is
a local space-time rapidity~\cite{HiranoTsuda}
which is shifted from the global midrapidity
according to the difference of the 
thickness of colliding nuclei
at $x_\perp$.
The CGC initial condition has no boost invariant region
due to the $x$ dependence of unintegrated gluon distribution
functions.
It is interesting to see the transverse profile from the CGC
and to compare with other parametrizations since
transverse presure gradient causes radial flow
in hydrodynamics.
Three initial conditions in the
transverse direction are compared
in Fig.~\ref{fig:ini_dens} (b).
In Ref.~\cite{HiranoTsuda}, transverse profile
for the initial energy density scales with
the number density of the binary collisions
(b.~c.~scaling).
On the other hand, the flat profile with smearing near the edge
of a nucleus is employed in Ref.~\cite{Morita}.
Note that, in Ref.~\cite{Morita}, the initial condition
is designed for $b=0$ fm collisions and that cylindrical
symmetry is explicitly assumed. 
For initial energy density taken from Ref.~\cite{Morita},
we take account of correction factors 1.1 and $(1.0/0.6)^{4/3}$
coming from the differences of collision energy
and initial time respectively.
The value of maximum energy density varies from 13 to 43
GeV/fm$^3$. However, these initial
conditions work at least for pion $p_T$ spectrum by adjusting thermal
freezeout temperature.

\section{hydrodynamics and parton energy loss}
\label{sec:hydrojet}

In this section, we summarize essential features of the
hydro+jet model~\cite{HN1,HN2,HN3,HN4}.

Once an initial condition in a
 hydrodynamic simulation is assigned
at an initial time $\tau_0$,
one describes the space-time evolution of thermodynamic
variables by numerically solving
hydrodynamic equations $\partial_\mu T^{\mu\nu} = 0$ with
the help of the EoS, $P = P(e)$. We are assuming
baryon free fluids here, so pressure $P$ is a function
of energy density $e$ only.
In the ideal hydrodynamics, the energy momentum tensor 
is $T^{\mu\nu} = (e+P) u^\mu u^\nu - Pg^{\mu\nu}$,
where $u^\mu$ is the four fluid velocity.
Assuming massless free parton system as discussed in the previous
section,
the EoS for the QGP phase is $P=(e-4B)/3$.
For the hadron phase,
we employ the partial chemical equilibrium model \cite{HiranoTsuda}
which describes the early chemical freezeout picture
by introducing chemical potential for each hadron.
Both the phase transition temperature
and the chemical freezeout temperature
are taken as $T_{\mathrm{c}} = T^{\mathrm{ch}} = 170$ MeV.
As a usual prescription, the bag constant $B= 486$ MeV/fm$^3$
is chosen by matching condition
for pressure
$P_{\mathrm{QGP}} (T_{\mathrm{c}})=P_{\mathrm{hadron}}(T_{\mathrm{c}})$. 
It should be emphasized that,
at collider energies, relativistic coordinate
in the time and longitudinal directions
$(\tau, \eta_{\mathrm{s}})$ is 
a substantial choice
to describe the evolution in the whole space-time
as discussed in Ref.~\cite{Hirano:2001eu,HiranoTsuda}.
Within hydrodynamic approaches, the studies of
centrality and rapidity dependences
of observables simultaneously
at the collider energies
are only accomplished by fully 3D
hydrodynamic simulations with the relativistic coordinate.
The statistical model and blast wave model analyses tell us
that the kinetic (thermal) freezeout temperature $T^{\mathrm{th}}$
is smaller than the chemical freezeout temperature
$T^{\mathrm{ch}}$ \cite{SH}.
Whereas, 
both thermal and chemical
equilibrium is assumed
in the conventional EoS.
This means that both freezeouts
happen simultaneously and that, consequently, one cannot reproduce
the $p_T$ spectra and the particle ratio 
at the same time.
By taking account of the chemical potential 
accompanied with the number conservation of
each stable hadron between chemical freezeout and kinetic freezeout,
we obtain the EoS denoting the
picture of early chemical freezeout.
Note that this EoS results in the almost $T^{\mathrm{th}}$ independent
$p_T$ slope for pions \cite{HiranoTsuda}.

For the hard part of the model,
  we generate momentum spectra of hard partons
 by  PYTHIA 6.2~\cite{pythia} in which
  $2\to2$ pQCD hard processes are included.
  Initial and final state radiations are
  taken into account for
 the enhancement of higher-order contributions
  associated with multiple small-angle parton emission.
EKS98 nuclear shadowing~\cite{eks98} is used
assuming the impact parameter dependence as in Ref.~\cite{Emel'yanov:1999bn}.
We employ the model in Ref.~\cite{XNWang:1998ww}
to take into account the multiple initial state scatterings,
in which initial $k_{T}$ is broadened proportional to the
number of scatterings.
Initial transverse positions of jets at an impact parameter $b$
 are determined randomly
 according to the probability specified by
 the number of binary collision distribution.
Jets are freely propagated up to the initial time $\tau_0$ of hydrodynamic
simulations
neglecting the possible interactions in the pre-thermalization stages.
Jets are assumed to travel with a straight line trajectory
in a time step.

Jets lose their energies through gluon emission induced
by the dense medium. A lot of work has already done by
many authors \cite{BDMPS,Zakharov,Wiedemann,GLV,WW,Arnold:2002ja}.
Here we employ the Gyulassy-Levai-Vitev (GLV)
formula~\cite{GLV} based on an opacity expansion approach
which is a relevant formalism for heavy ion collisions.
In the original formula,
the amount of energy loss
is expanded into opacity.
Here we take the first order term and
neglect the kinematics of emitted gluons
\begin{equation}
\Delta E = C \int_{\tau_0}^{\infty} d\tau
\rho\left(\tau, \bm{x}\left(\tau\right)\right)
(\tau-\tau_0)\ln\left({\frac{2p_0 \cdot u}{\mu^2 L}}\right),
\label{eq:GLV}
\end{equation}
where $C$ includes the strong running coupling and color Casimir factors and
$\rho$ is a %thermalized
 parton density which is to be taken from full 3D
hydrodynamic simulations in our approach.
$\mu$(=0.5GeV) is a screening mass and
$L$(=3fm) is a typical length of a medium which can be identified
with a lifetime of partonic phase. $\bm{x}$ and $p_0$ are
a position and an initial four momentum of a jet.
The existence of $\tau-\tau_0$ comes from the so-called
LPM effect~\cite{LPM}.
In the static medium case, $p_0 \cdot u$ is replaced by 
an initial energy of a jet $E_0$.

In the actual simulations, the coefficient $C$
in Eq.~(\ref{eq:GLV}) is regarded as an adjustable
parameter roughly corresponding to an 
averaged value over quarks and gluons up to some factor.
Our strategy here is that
a constant $C$ is set by fitting the PHENIX data on $R_{AA}$
for neutral pions in the most central events at midrapidity
\cite{phenix:pi0} and
that we next look at the centrality and rapidity
dependences.
The main reasons are the following.
First, we neglect fluctuations of the radiated gluon number~\cite{Baier:2001yt}
in our calculations.
An absolute value of the energy loss is about factor two smaller
when one takes into account the fluctuations of the radiated gluon number
at RHIC~\cite{GLV2}.
However, 
the $p_T$ dependence of $R_{AA}$ is found
to be the same at RHIC energies~\cite{GLV2}.
Second, in our present hydrodynamic calculations,
it is assumed that massless free gas for quarks ($N_f=3$) and gluons.
However,
it is commonly believed that gluons are 
dominant components in the early
stages of collisions
at collider energies because of the gluon dominance of
the parton structure function at small-$x$.
Thus inclusion of specie dependent factors in $C$
may lead to discrepancy between our hydro+jet results and data.

\section{Results}
\label{sec:result}

In this section, we first study the centrality dependence of 
longitudinal and transverse distribution.
Next, the results for
the centrality dependence of
 the nuclear modification factors
and back-to-back correlations are presented.
Finally, we study the nuclear modification factor
 at forward rapidity.

Let us summarize our model parameters before showing the results.
In the following, we use
the parameters $K=0.7$ and $\lambda=0.2$
in the gluon distribution of a nucleon.
$\kappa$ depends on how
to match initial distributions.
For IC-$e$ (IC-$n$),
$\kappa^2 = 2.25 (3.6)$.
In the hydrodynamic calculation,
the kinetic freezeout temperature of $T^{\mathrm{th}}=100$ MeV
and the initial time of $\tau_0 = 0.6$ fm/$c$
are taken. We will not analyze the detailed $T^{\mathrm{th}}$ 
dependence, but will make a short comment on that.
A parameter $C=0.35$ in the parton energy loss formula
is fixed by fitting the nuclear modification factor for neutral
pions from PHENIX data at 10\% centrality~\cite{phenix:pi0}.

\subsection{Rapidity, centrality, and energy dependences of hadrons}
\label{sec:dndy}

%%%%%%%%%%%%%%%%%%%%%%%%%%%%%%%%%%%%%%%%%%%%%%%%%%%%%%%%%%%%
% hydro rapidity distribution
%%%%%%%%%%%%%%%%%%%%%%%%%%%%%%%%%%%%%%%%%%%%%%%%%%%%%%%%%%%%
\begin{figure}[t]
\includegraphics[width=3.3in]{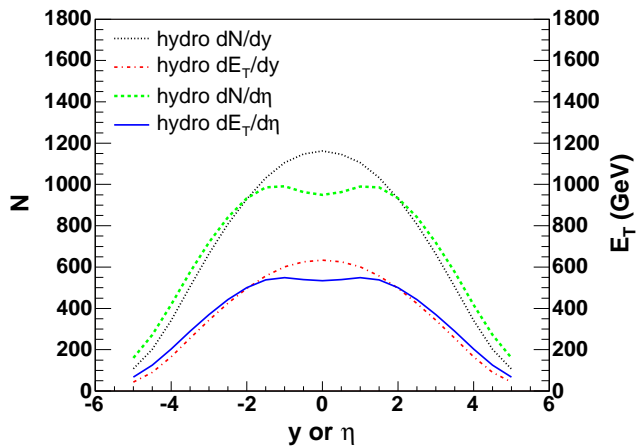}
\caption{
Rapidity and pseudorapidity distributions of all hadrons
 in Au + Au  collisions at $\sqrt{s_{NN}}=200$ GeV at $b=2.4$ fm
are compared to the corresponding transverse energy distributions
obtained from hydrodynamic simulations with CGC initial condition.
}
\label{fig:dndy_final}
\end{figure}
%%%%%%%%%%%%%%%%%%%%%%%%%%%%%%%%%%%%%%%%%%%%%%%%%%%%%%%%%%%%

(Pseudo)rapidity distributions of all hadrons
and transverse energy in central Au+Au collisions
from the hydrodynamic simulation
are shown in Fig.~\ref{fig:dndy_final}.
Here IC-$n$ is used.
As one can compare Fig.~\ref{fig:dndy_final} with Fig.~\ref{fig:ini_dens} (a),
the effect of the hydrodynamic afterburner is to reduce
the transverse energy per particle
due to $pdV$ work.
We find that $(dE_T/dy)/(dN/dy)|_{y=0} = 0.54$ GeV
and $(dE_T/d\eta)/(dN/d\eta)|_{\eta=0}=0.56$ GeV.
Recall that our CGC initial condition from Eq.~(\ref{eq:ktfac})
yields $E_T/N_g=1.6$ GeV
and that it becomes $E_T/N_g=1.0$ GeV after assuming a thermal state 
from Fig.~\ref{fig:ini_dens} (a).
The consequence of the hydrodynamic evolution on the shape
of rapidity distribution is to make it wider very slightly.
Final shape of the rapidity distribution
looks like Gaussian which is consistent with recent
data from BRAHMS~\cite{brahms:dndy}.
Our result supports that the 
KLN calculation~\cite{KLN}
which is based on the assumption of parton-hadron
duality is a good approximation for the discussion of
the shape of pseudorapidity distribution.
The number of initial gluons at midrapidity is $\sim 1480$
(see Fig.~\ref{fig:ini_dens} (b)),
which can be comparable with
the number of final hadrons at midrapidity ($\sim 1160$).
One see in Fig.~\ref{fig:dndy_final} that pseudorapidity distributions
for both number and transverse energy are almost flat at mid-rapidity
in the range of $|\eta|<2$, although rapidity distribution has no
flat region.

%%%%%%%%%%%%%%%%%%%%%%%%%%%%%%%%%%%%%%%%%%%%%%%%%%%%%%%%%%%%
% hydro psedorapidity distribution at 130/200 GeV
%%%%%%%%%%%%%%%%%%%%%%%%%%%%%%%%%%%%%%%%%%%%%%%%%%%%%%%%%%%%
\begin{figure*}[ht]
\includegraphics[width=3.5in]{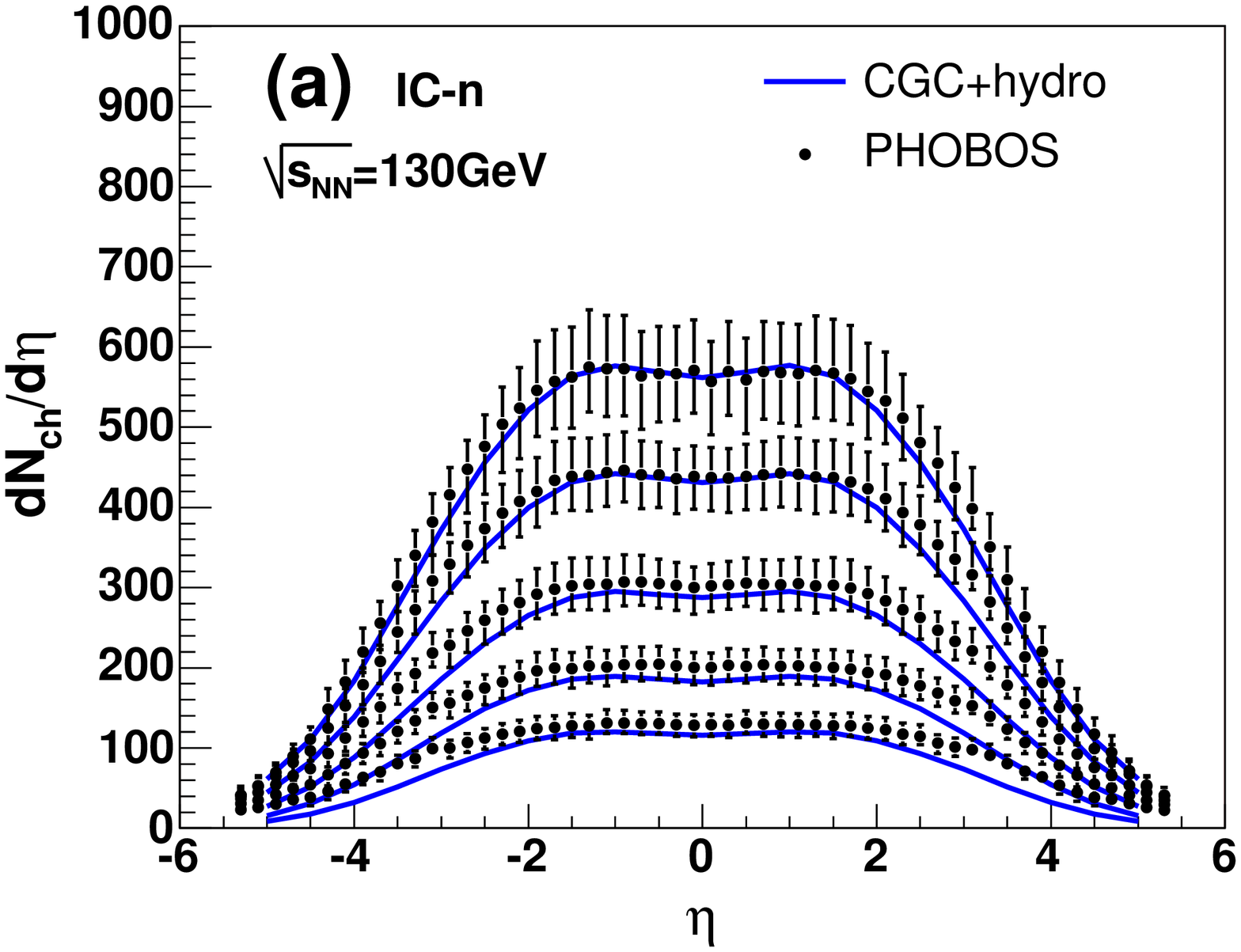}
\includegraphics[width=3.5in]{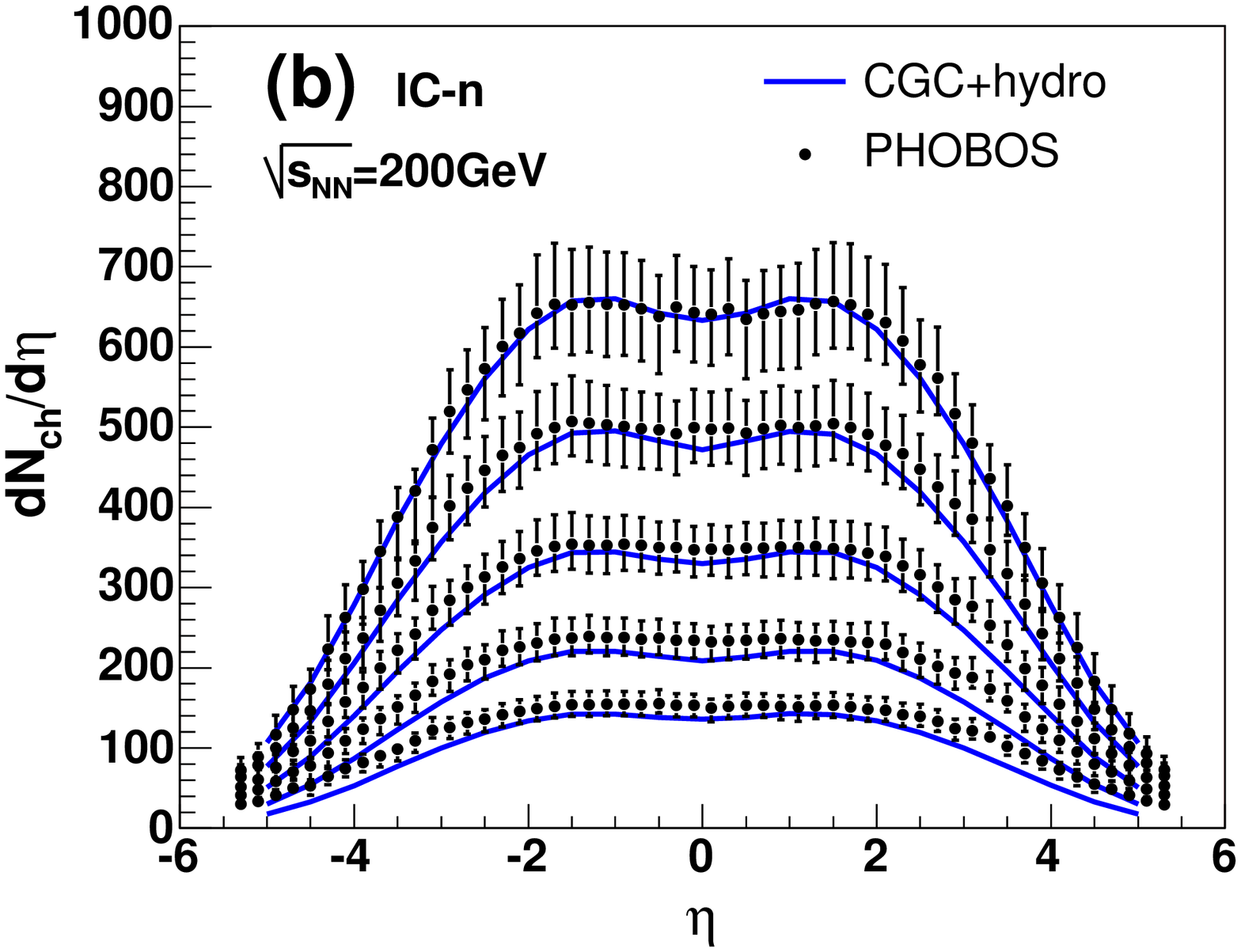}
\includegraphics[width=3.5in]{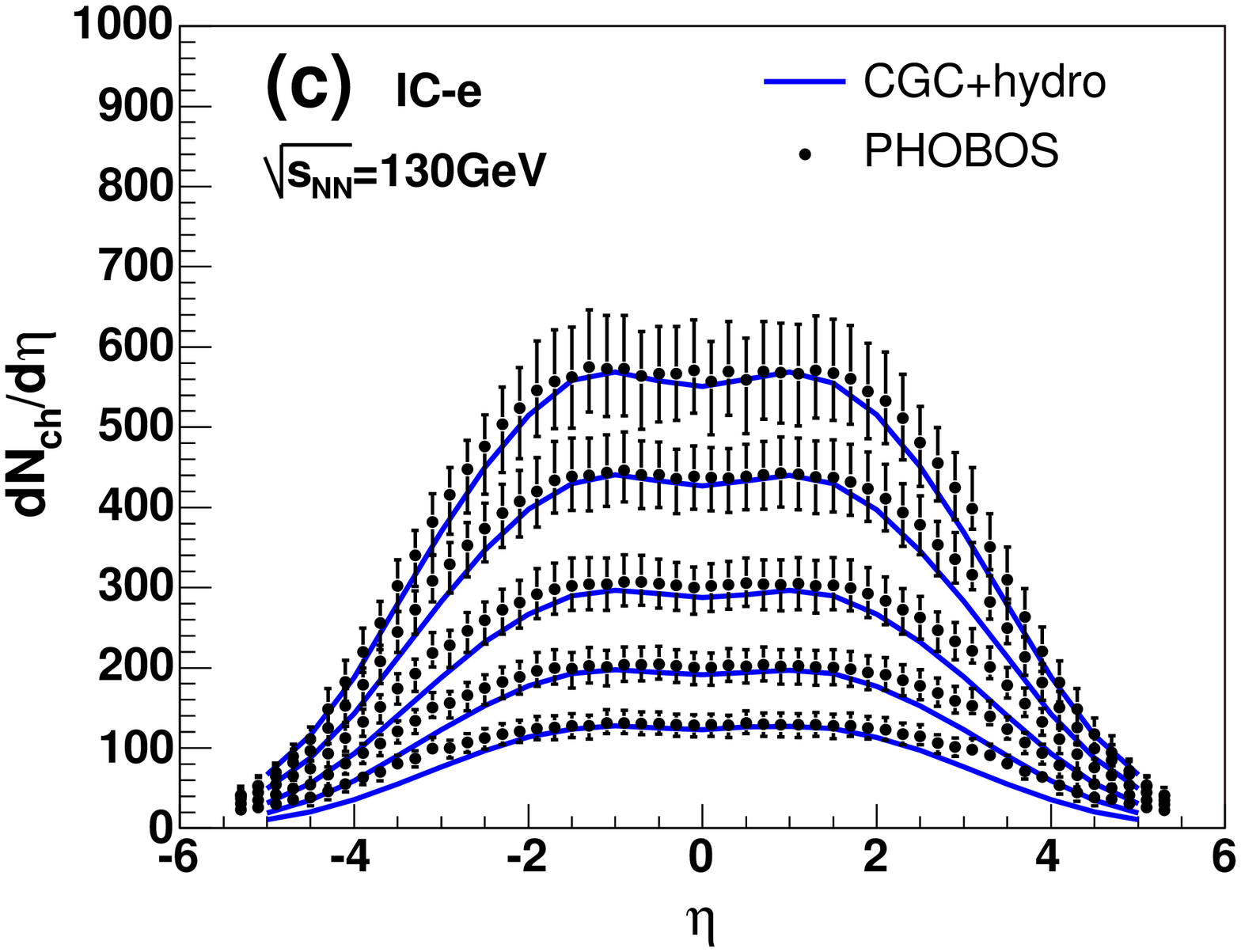}
\includegraphics[width=3.5in]{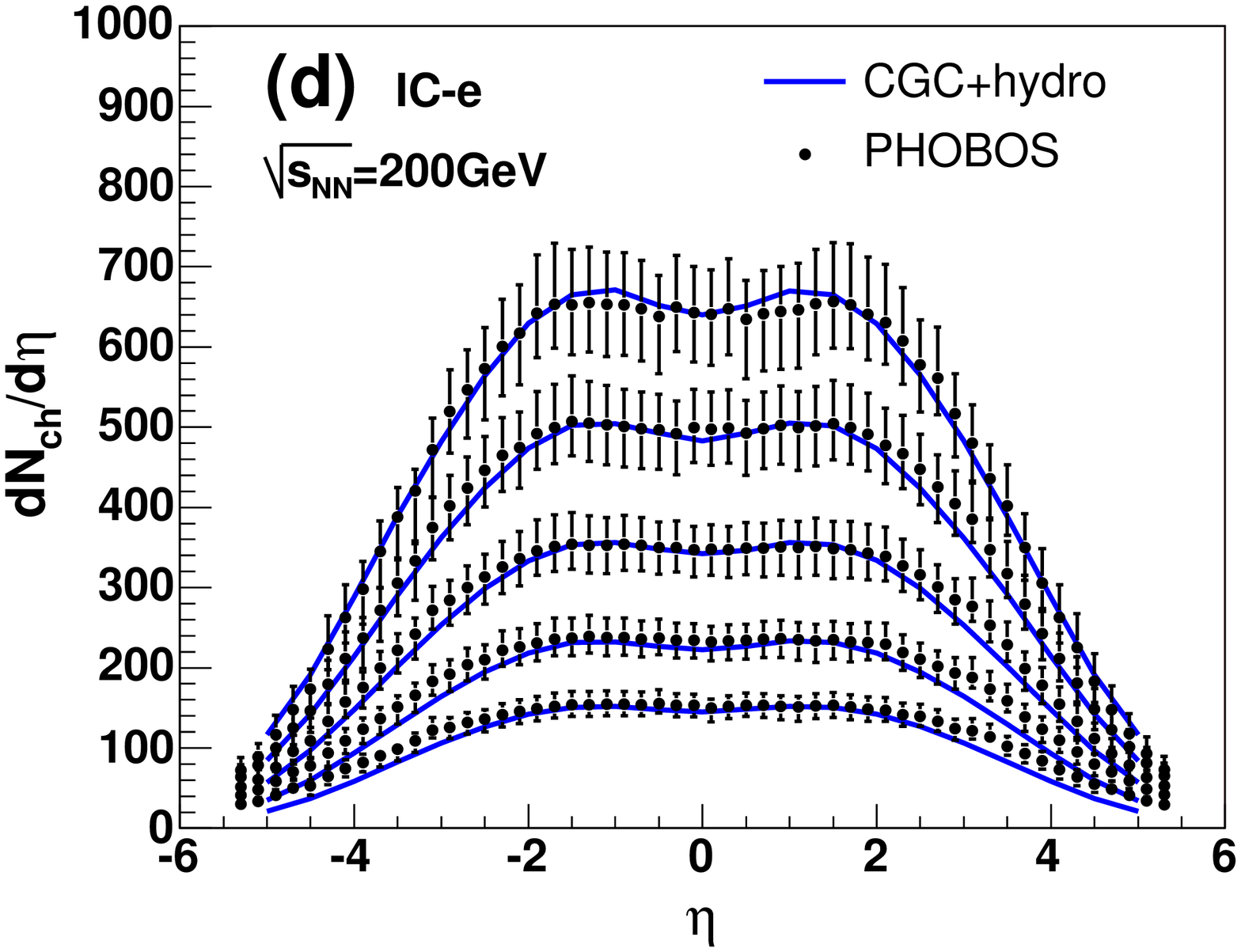}
\caption{
Pseudorapidity distributions of charged hadrons
 in Au + Au  collisions at $\sqrt{s_{NN}}=130$ and 200 GeV
are compared to the PHOBOS data~\cite{Back:2002wb}.
For the choice of initial conditions and impact parameters, see text.
}
\label{fig:dndeta200}
\end{figure*}
%%%%%%%%%%%%%%%%%%%%%%%%%%%%%%%%%%%%%%%%%%%%%%%%%%%%%%%%%%%%

The main difference between the previous~\cite{Hirano:2001eu,HiranoTsuda}
and the present initial conditions
is whether boost invariant region exists or not
(see Fig.~\ref{fig:ini_dens} (a)).
One may ask why the previous initial conditions with boost invariant
region near midrapidity 
can also result in the Gaussian shape in rapidity distributions.
It would be instructive here to discuss about the boost invariance for
thermodynamic quantities and particle distributions
within a hydrodynamic framework.
Non-boost invariant $dN/dy$
recently observed by BRAHMS~\cite{brahms:dndy}
reminds us the so-called Landau picture~\cite{Landauhydro}.
Boost invariance for $dN/dy$ is equivalent to the boost invariance
of thermodynamic quantities
in the original Bjorken assumption~\cite{Bjorken:1982qr} 
in which the length in the $\eta_s$ direction
is assumed to be infinite.
However, this is not always true
if boost-invariant region for thermodynamic variables is finite.
In such a case, rapidity distribution can have
\textit{no} boost-invariant region.
Let us consider a simple example.
Rapidity distribution can be related to the phase-space distribution
function via Cooper-Frye formula~\cite{Cooper:1974mv}:
\begin{equation}
   \frac{dN}{dy} \propto \int d^2p_T \int f(x,p)p^{\mu}d\sigma_{\mu}.
\end{equation}
Consider, e.g., a 
finite Bjorken cylinder
$\mid\eta_s\mid < \eta_0$
with a radius $R$
for massless pions. 
Assuming no transverse flow,
we have
 $\int d\sigma^{\mu}=\int
d\eta_s \pi R^2 \tau (\cosh\eta_s,\bm{0}_\perp,\sinh\eta_s)$
in the case of flat temperature profile.
Thus the rapidity distribution becomes
\begin{eqnarray}
 \frac{dN}{dy} &\propto& \int^{\eta_0}_{-\eta_0}d\eta_s
                \int^{\infty}_{0} dp_T p_T^2\cosh(\eta_s-y)
              f(\vec{p}) \nonumber\\
  &\propto& \int^{\eta_0}_{-\eta_0}d\eta_s
        \frac{T^3}{\cosh^2(\eta_s-y)},
\end{eqnarray}
where $f(\vec{p}) = 1/(\exp[p_T\cosh(\eta_s-y)/T] -1)$.
This shows rapidity distribution for massless pions
in a fluid element at $\eta_{\mathrm{s}}$ 
becomes $\propto 1/\cosh^2(y-\eta_{\mathrm{s}})$.
After the above distribution is folded in the finite
boost-invariant region $\mid\eta_s\mid<\eta_0$,
one obtain
\begin{equation}
  \frac{dN}{dy} \propto  T^3 [\tanh(y+\eta_0) - \tanh(y-\eta_0)].
\end{equation}
The resultant rapidity distribution still looks like
a Gaussian shape when $\eta_0 < 2$.
The flat region starts to 
appear
when $\eta_0 \gsim 4$.
For more detailed discussions including finite mass effects,
see Ref.~\cite{Schnedermann:1993ws}.
We actually reproduced
the PHOBOS ($\sqrt{s_{NN}}=$130 GeV) 
and the BRAHMS ($\sqrt{s_{NN}}=$200 GeV)
data by using such a
Bjorken like initial condition
 in Ref~\cite{Hirano:2001eu,HiranoTsuda,HN4}
(for an initial condition,
see also Fig.~\ref{fig:ini_dens} (a)).
%%%%%%%%%%%%%%%%%%%%%%%%%%%%%%%%%%%%%%%%%%%%%%%%%%%%%%%%%%%%%%%%%%%%%%%%%%%
% initial dEt/dy in  CGC
%%%%%%%%%%%%%%%%%%%%%%%%%%%%%%%%%%%%%%%%%%%%%%%%%%%%%%%%%%%%%%%%%%%%%%%%%%%
\begin{figure}[t]
\includegraphics[width=3.5in]{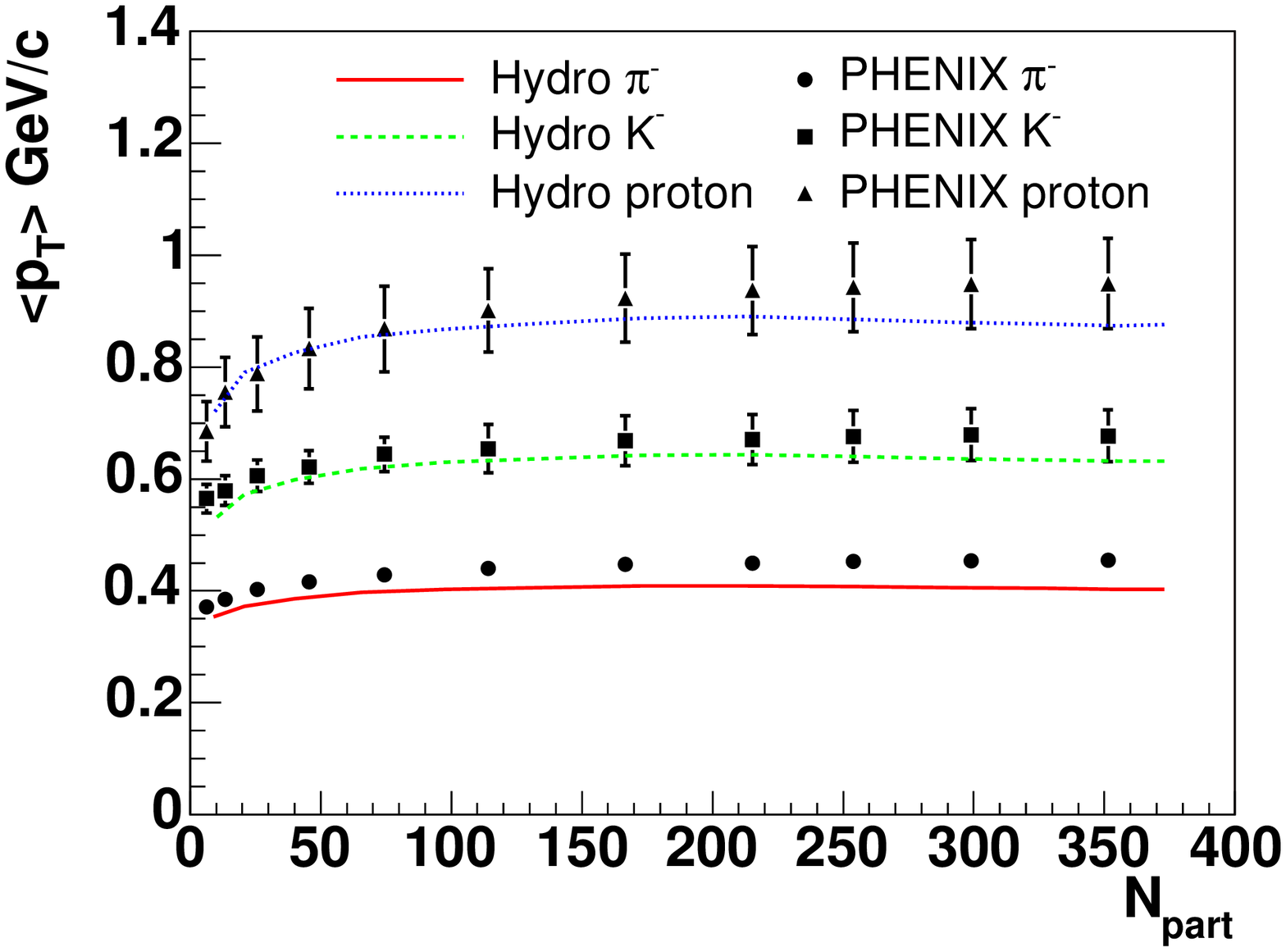}
\caption{
Mean transverse momenta for pions, kaons and protons
as a function of $N_{\mathrm{part}}$.
Here contribution only from hydrodynamic components is taken into account.
$T^{\mathrm{th}}=100$ MeV is used for all centralities.
Data are taken from Ref.~\cite{phenix:pi}
}
\label{fig:meanpt}
\end{figure}
%%%%%%%%%%%%%%%%%%%%%%%%%%%%%%%%%%%%%%%%%%%%%%%%%%%%%%%%%%%%

%%%%%%%%%%%%%%%%%%%%%%%%%%%%%%%%%%%%%%%%%%%%%%%%%%%%%%%%%%%%
% hydro pT distribution
%%%%%%%%%%%%%%%%%%%%%%%%%%%%%%%%%%%%%%%%%%%%%%%%%%%%%%%%%%%%
\begin{figure*}[t]
\includegraphics[width=2.3in]{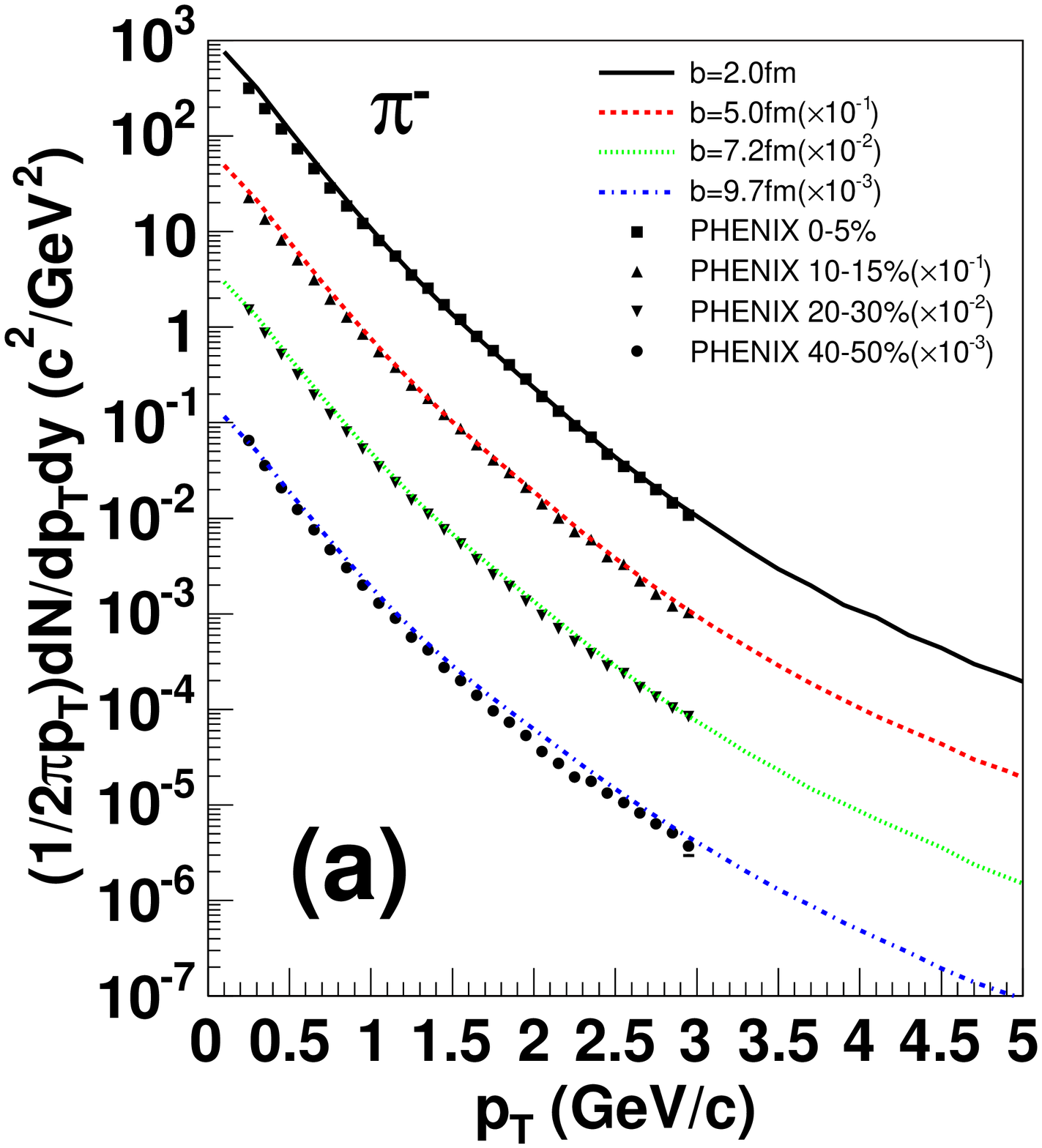}
\includegraphics[width=2.3in]{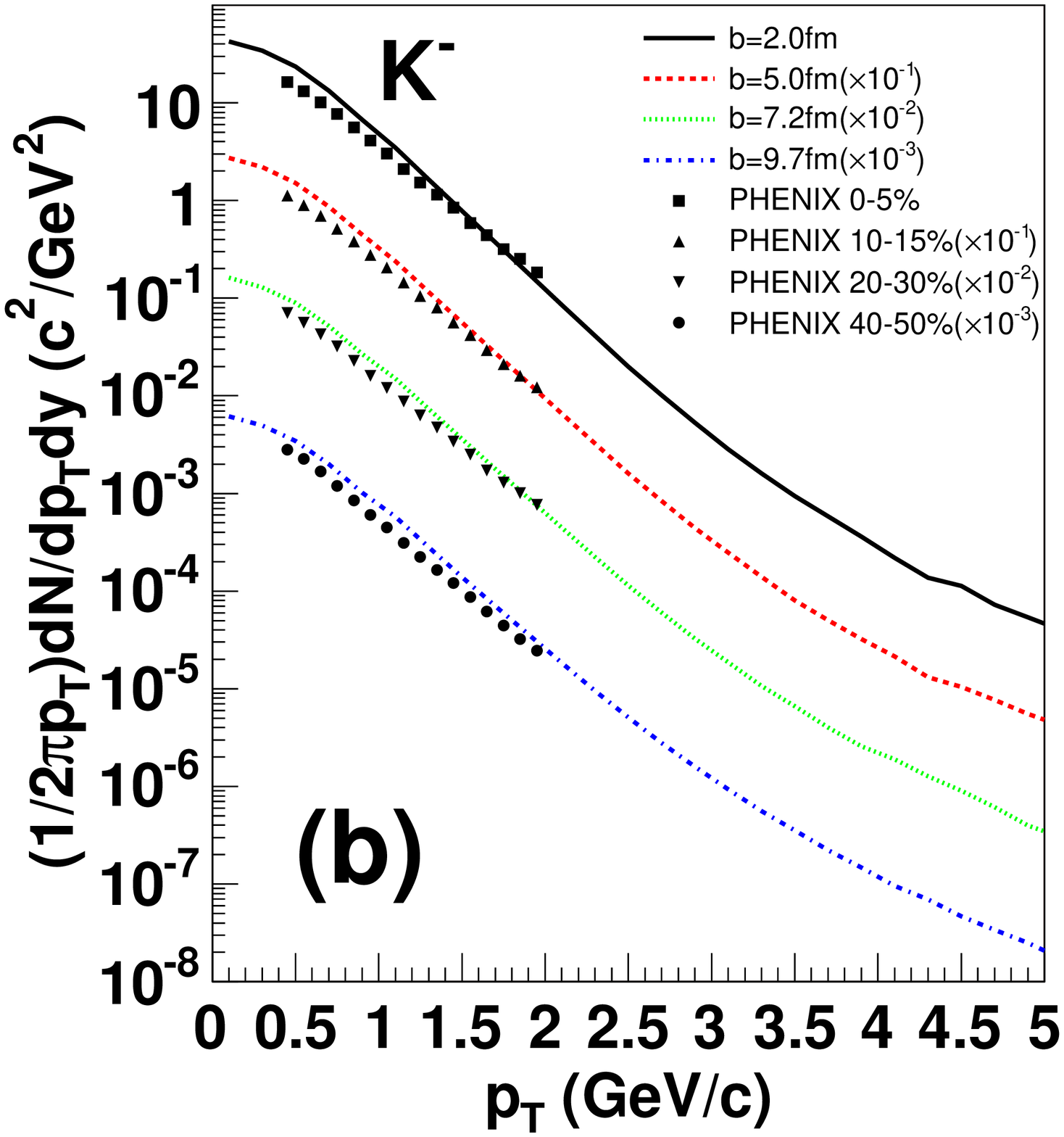}
\includegraphics[width=2.3in]{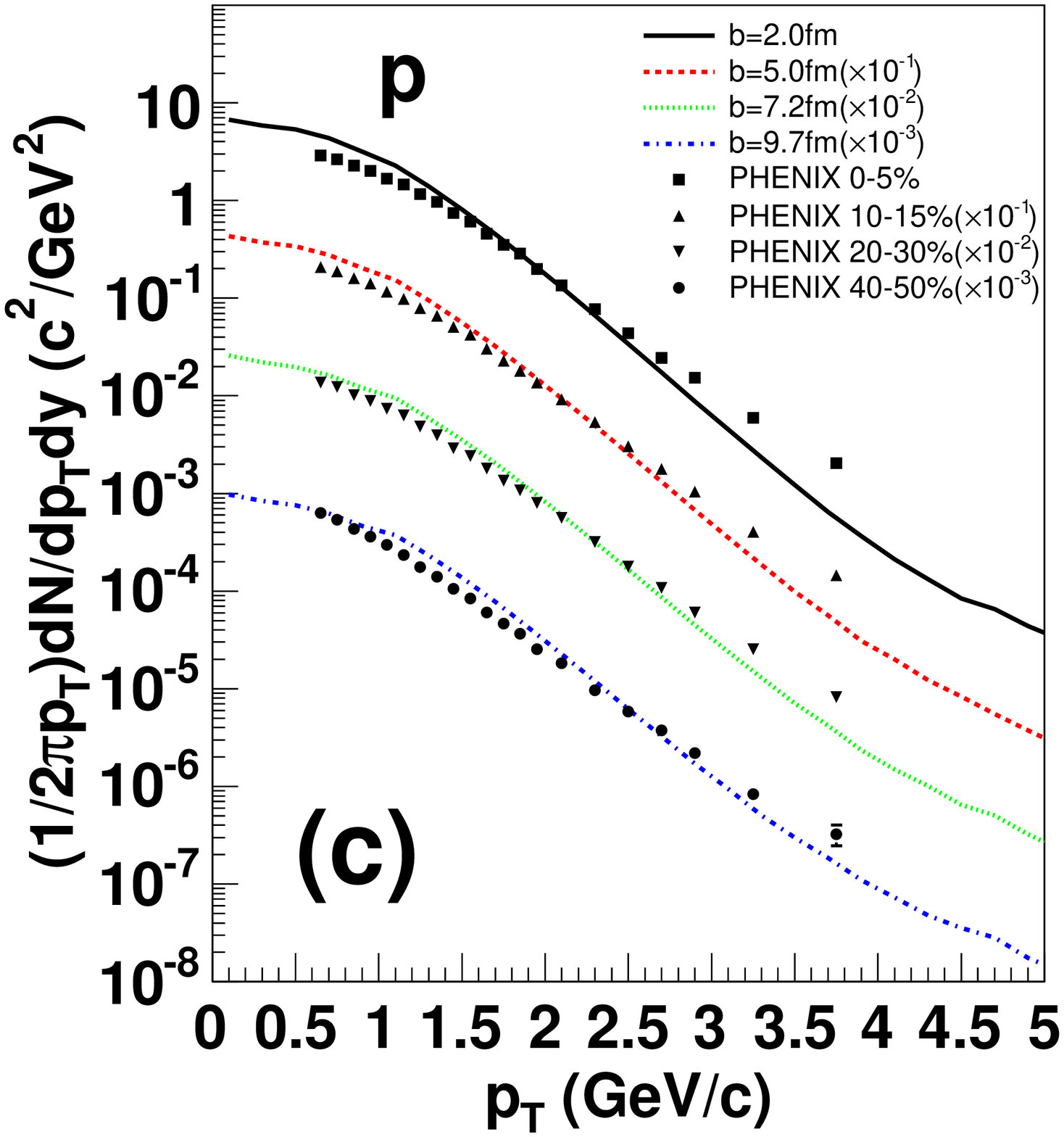}
\caption{
Centrality dependences of the $p_T$ spectra
for (a) pions, (b) kaons, and (c) protons
obtained from the CGC+hydrojet model
are compared with the PHENIX data \cite{phenix:pi}.
Kinetic freeze-out temperature $T^{\mathrm{th}}=100$
MeV is used in the calculations,
}
\label{fig:dndpt}
\end{figure*}
%%%%%%%%%%%%%%%%%%%%%%%%%%%%%%%%%%%%%%%%%%%%%%%%%%%%%%%%%%%%

In Fig.~\ref{fig:dndeta200}, pseudorapidity distributions
of charged hadrons in Au + Au collisions at both $\sqrt{s_{NN}}=130$
and 200 GeV are compared with the PHOBOS data~\cite{Back:2002wb}.
Figures 4 (a) and (b) (Figures 4 (c) and (d))
show the results of the initial condition
which matches
the number (energy) density distributions for partons
between the CGC and hydrodynamics.
Impact parameters in each panel
are, from top to bottom,
2.4, 4.5, 6.3, 7.9, and 9.1 fm
(2.5, 4.4, 6.4, 7.9, and 9.1 fm)
for $\sqrt{s_{NN}} =$ 200 (130) GeV.
These impact parameters
are evaluated from
the average number of participants
at each centrality
estimated by PHOBOS~\cite{Back:2002wb}.
We are not interested in fit for all rapidity region, because
we do not include baryon chemical potentials and
both the CGC and hydrodynamics break down when density becomes small.
The results within the range 
$|\eta| \lsim 3$-4 are satisfactory for reproduction of data.
Therefore, KLN $k_T$ factorization approach in the CGC provides
very good initial conditions
for the hydrodynamic simulations which reproduce rapidity,
centrality and energy dependences.
It should be also emphasized that it is not easy to parametrize
such initial conditions which reproduce the data with the same
quality as the CGC initial conditions presented here.

In the calculations, we use $\lambda=0.2$ to get the best fit.
However, we find that the results with the range of $\lambda=0.2$-0.3
are still within experimental error bars.
Note that a good description of HERA data
has been obtained by parametrization of
the saturation scale $Q^2_s \sim x^\lambda$
with $\lambda \simeq 0.25$-0.3~\cite{GBW,IIMu03}.

Kinetic freeze-out temperature $T^{\mathrm{th}}$ which 
is conventionally
fixed by the slope of the transverse momentum spectra
is a free parameter in hydrodynamics.
However, $p_T$ slope for pions becomes insensitive to $T^{\mathrm{th}}$
when one considers early chemical freezeout \cite{HiranoTsuda,Kolb:2002ve}.
But, we need to check the $T^{\mathrm{th}}$ dependence on the multiplicity.
When we change the kinetic freeze-out temperature,
we can also reproduce
the data by
$\kappa^2 $ = 3.5, 3.6, 3.7, and 3.88
%3.3, 3.56, 3.64, and 3.8
for $T^{\mathrm{th}}$ = 80, 100, 120, and 140 MeV, respectively,
for IC-$n$.
We also investigate the $\tau_0$
dependence on the particle multiplicity.
We find that the particle yield
with $\tau_0=1.0$ fm/$c$ is only 4\% larger
than the case $\tau_0=0.6$ fm/$c$
regardless of the centralities as well as energies.
This difference can be understood by the $pdV$ work effect.

Mean transverse momenta $\langle p_T \rangle$
for pions, kaons and protons
as a function of $N_\mathrm{part}$ 
are compared with the PHENIX data \cite{phenix:pi}
in Fig.~\ref{fig:meanpt}.
Since the mean transverse momenta reflects
the low $p_T$ physics especially radial flow, contribution from
hydrodynamic components at midrapidity is considered in this calculations.
Although our results are slightly smaller than the data
in central and semicentral regions, the overall trend is consistent
with the data.
For pions, the semihard spectrum starts to be comparable with the soft
spectrum around $p_T=1.5$-2 GeV/$c$ in this approach.
As we will see in the next subsection, we reproduce the $p_T$ spectra
for pions by including contribution from jets.
So the deviation for pions can be filled by the semihard spectrum.
While semihard components for kaons and protons are very small
in low and intermediate $p_T$ regions.
A little more radial flow
is needed to gain the mean transverse momentum
in central collisions.
More detail discussion will be given in the next subsection.

%%%%%%%%%%%%%%%%%%%%%%%%%%%%%%%%%%%%%%%%%%%%%%%%%%%%%%%%%%%%%%%%%%%%%%%%%%%
% initial dEt/dy in  CGC
%%%%%%%%%%%%%%%%%%%%%%%%%%%%%%%%%%%%%%%%%%%%%%%%%%%%%%%%%%%%%%%%%%%%%%%%%%%
%\begin{figure}[ht]
%\includegraphics[width=3.5in]{dndeta200eden.eps}
%\caption{
%Comparesion between the eCGC and nCGC on the final 
% pseudorapidity distribution of charged hadrons
% in Au + Au  collisions at $\sqrt{s_{NN}}=200$ GeV.
%}
%\label{fig:dndeta200eden}
%\end{figure}
%%%%%%%%%%%%%%%%%%%%%%%%%%%%%%%%%%%%%%%%%%%%%%%%%%%%%%%%%%%%

\subsection{Jet quenching}

In this subsection, we compute the centrality and rapidity dependences
of high $p_T$ spectra and nuclear modification factors
in the forward rapidity region.
All results are obtained by using the initial condition IC-$n$.
In our model, high $p_T$ jets suffer interaction
with the local parton density whose evolution
is governed by hydrodynamics with the CGC initial conditions.
We only take into account parton
energy loss in the deconfined matter $T \ge T_c$.

Figure \ref{fig:dndpt} shows that the $p_T$ spectra
for identified hadrons from the CGC+hydrojet model
 are compared with the PHENIX data~\cite{phenix:pi}
for several centrality bins.
The final $p_T$ spectrum in our approach is the sum of
the contribution from fluid elements and from fragmentation of jets.
First of all, we see that spectra obtained with the CGC initial condition
yield the similar results with the initial condition
employed previously~\cite{HN4}.
Agreement for the pion spectra with the PHENIX data is perfect.
As mentioned before, hydrodynamic component is
insensitive to the choice of $T^{\mathrm{th}}$ for the pion spectrum
in the chemically frozen hydrodynamics.
Slopes of kaons and protons in the model prediction
are steeper than that of the data.
This indicates that
more radial flow is needed 
for kaons and protons to
reduce yields in the low $p_T$ region
and to enhance in the intermediate $p_T$ region simultaneously,
even after inclusion of the jet components into hydrodynamics.
We have checked that
freeze-out temperature of $T^{\mathrm{th}}=80$ MeV
still underestimates the slope.
RHIC data suggest that
initial transverse flow at the time of thermalization $\tau_0$
seems to be important
as studied by Kolb and Rapp~\cite{Kolb:2002ve}.
This observation is natural because
one expects that many rescatterings lead to thermalization.
Initial flow profile
can be calculated from the classical
Yang-Mills fields.
Therefore, implementation of Yang-Mills results
into the initial condition in the hydrodynamic simulations
could considerably improve (modify) the hydro results on the transverse
momentum distributions.
Discrepancy in the high $p_T$ for the proton spectrum
might be improved by considering the recombination among
hard partons and soft partons~\cite{Hwa:2004ng}.

\begin{figure}[ht]
\includegraphics[width=3.5in]{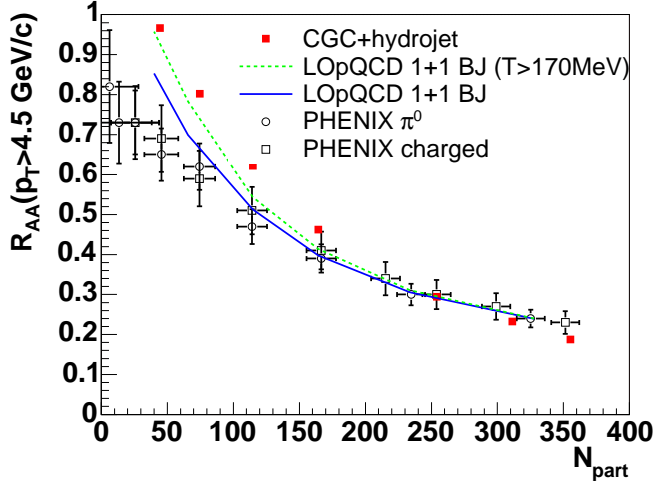}
\caption{
The CGC+hydrojet result on
nuclear modification factor $R_{AA}(p_T>4.5\mathrm{GeV/c})$
 in Au + Au  collisions at $\sqrt{s_{NN}}=200$
is compared with the PHENIX data \cite{Adler:2003au}.
}
\label{fig:RAA45}
\end{figure}

In Fig.~\ref{fig:RAA45}, we investigate
the centrality dependence of the nuclear modification factor
for $|\eta|<0.35$
\begin{equation}
  R_{AA}(p_T> 4.5\mathrm{GeV/}c) = \frac{\int_{4.5\mathrm{GeV/}c}dN^{AA}}
{N_{\mathrm{coll}}\int_{4.5\mathrm{GeV/}c}dN^{pp}}
\end{equation}
together with the PHENIX data \cite{Adler:2003au}.
Our results only account for the data up to mid-central events
and fail to reproduce the data in peripheral collisions.
Although this is reasonable because neither CGC nor hydrodynamic description
can be reliable in small particle densities,
let us study more detailed mechanism.

%%%%%%%%%%%%%%%%%%%%%%%%%%%%%%%%%%%%%%%%%%%%%%%%%%%%%%%%%%%%
% RAA pt > 4.5GeV
%%%%%%%%%%%%%%%%%%%%%%%%%%%%%%%%%%%%%%%%%%%%%%%%%%%%%%%%%%%%
%%%%%%%%%%%%%%%%%%%%%%%%%%%%%%%%%%%%%%%%%%%%%%%%%%%%%%%%%%%%

In the recent studies~\cite{Wang:2003mm,Wang:2003aw,Drees},
it is claimed that
the centrality dependence of single and dihadron suppression
spectra even in peripheral collisions is described by jet quenching
in the partonic phase.
They assume in their calculation that
the parton density is scaled with the number of
participants and
that the system expands according to
(1+1)-dimensional Bjorken
expansion \cite{Bjorken:1982qr}.
Line integral for the loss of parton energy
is evaluated at all densities.
This means that there exists only the partonic phase
even in very small partonic density
and that the effects of phase transition
or the hadronic phase are neglected
in those calculations.
In order to study discrepancy
between their results and our results
in peripheral collisions,
we also estimate the centrality dependence
of $R_{AA}$ assuming (1+1)-dimensional Bjorken expansion
with the same initial conditions
as the 3D ones.
In this calculation, 
the transverse shape of the initial condition
 is unchanged and the magnitude
of the parton density decreases as
$\rho(\tau, \bm{x}_\perp) = (\tau_0/\tau)\rho(\tau_0, \bm{x}_\perp)$.
Note that our transverse profile scales as
$\rho \approx Q_s^2(b)/\alpha_s(Q^2_s(b)) 
   \approx \rho_{\text{part}}/\alpha_s(Q^2_s(b))$
at mid-rapidity.
%Hydrodynamic
pQCD calculations with (1+1)-D Bjorken expansion are performed 
until the maximum partonic temperature reaches $T=100$ MeV.
When jet quenching happens in $T \ge T_c$,
the result of this (1+1)-D Bjorken calculation %with limiting termperature
yields almost the same results as our full 3-D calculations
as plotted in the dotted line in Fig.~\ref{fig:RAA45}.
However, when we do not impose a restriction on
temperature in the parton
energy loss as $T \ge T_c$,
the result becomes closer to the data
(see solid line in Fig.~\ref{fig:RAA45}).
The difference between these two calculations
suggests that there exists a contribution to energy loss 
besides the purely partonic phase.
This can be understood as follows.
The density of the system essentially  decreases 
like $\sim 1/\tau$
even in our 3D hydrodynamics.
In the case of most central collisions,
initial density is so high
and most energy loss is likely to occur in the early stages of the
collisions.
On the other hand, initial energy loss
alone is not enough to lose energy of each parton
due to small life time of
the partonic phase in peripheral collisions.
Hence energy loss should last longer until the density
of the system becomes small.
From above considerations, hadronic interactions~\cite{Gallmeister:2002us}
seem to be required as well
in order to understand the suppression
in peripheral collisions, when the bulk matter
scales as the $dN/dy \sim N_{\mathrm{part}}/\alpha_s(Q_s^2)$
which is consistent with the centrality dependence of
the experimental data
contrast to the number of participants scaling.

%%%%%%%%%%%%%%%%%%%%%%%%%%%%%%%%%%%%%%%%%%%%%%%%%%%%%%%%%%%%
% I_{AA}
%%%%%%%%%%%%%%%%%%%%%%%%%%%%%%%%%%%%%%%%%%%%%%%%%%%%%%%%%%%%
\begin{figure}[ht]
\includegraphics[width=3.5in]{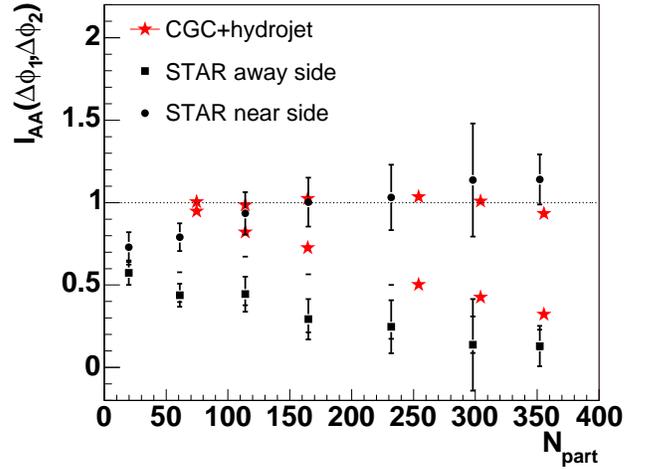}
\caption{
The CGC+hydrojet result on
$I_{AA}$ in Au + Au  collisions at $\sqrt{s_{NN}}=200$
is compared with the STAR data \cite{STAR:btob}.
}
\label{fig:IAA}
\end{figure}
%%%%%%%%%%%%%%%%%%%%%%%%%%%%%%%%%%%%%%%%%%%%%%%%%%%%%%%%%%%%

In Fig.~\ref{fig:IAA}, we plot the ratio of the strength
of correlation
in Au+Au and in $pp$ collisions
\begin{equation}
I_{AA}= \frac{\int^{\Delta\phi_2}_{\Delta\phi_1}d(\Delta\phi)
               C_2^{AuAu}(\Delta\phi)}
         {\int^{\Delta\phi_2}_{\Delta\phi_1}d(\Delta\phi)C_2^{pp}(\Delta\phi)}
\end{equation}
as a function of the number of participants $N_{\mathrm{part}}$
together with the experimental data from STAR~\cite{STAR:btob}.
$C_2(\Delta\phi)$ is the azimuthal pair distribution per trigger particle,
\begin{equation}
C_2(\Delta \phi) = \frac{1}{N_{\mathrm{trigger}}}
\int_{-1.4}^{1.4} d\Delta\eta
\frac{dN}{d\Delta\phi d\Delta\eta}.
\end{equation}
Here $\Delta \phi$ and $\Delta \eta$ are, respectively,
the relative azimuthal angle and pseudorapidity between a trigger
particle and an associated particle.
Charged hadrons in
$4 < p_{T,\mathrm{trigger}} < 6$ GeV/$c$
and in $2 < p_{T,\mathrm{associate}} < p_{T,\mathrm{trigger}}$
GeV/$c$ are defined to be trigger particles and associated particles
respectively.
In the calculation of $C_2(\Delta\phi)$, back ground is subtracted
as in Ref.~\cite{HN2}.
Near side correlation is defined as $\mid\Delta\phi\mid<0.75$ radian
and away side as $\mid\Delta\phi\mid>2.24$ radian.
Our results for the near side correlation show $I_{AA} \sim 1$
for all centralities, because
interactions of jets after their hadronization are not included.
However the data show $I_{AA}<1$ for the near side
in peripheral collisions.
The results on the away side correlation for $N_{\mathrm{part}}\lsim 100$
deviate from the data, which is consistent with the behavior of
$R_{AA}$ as discussed in Fig.~\ref{fig:RAA45}.

\subsection{Nuclear modification factors in the forward rapidity}

Recently, 
the pseudorapidity dependence of 
nuclear modification factors in both Au+Au and $d$+Au
collisions
observed by BRAHMS \cite{BRAHMS:dA,BRAHMS:qm2004}
attracts many theoretical interests
in the context of the saturation physics,
since one can go deeply into the small $x$ region
at forward/backward rapidity.
Previously, we investigated the nuclear modification factor
in Au+Au collisions
\begin{equation}
 R_\eta = \frac{R_{AA}(\eta=2.2)}{R_{AA}(\eta=0)}
\label{eq:reta}
\end{equation}
within the hydro+jet model~\cite{HN3}.
In Ref.~\cite{HN3}, rapidity dependence of
initial parton density was
assumed to be flat in $\mid \eta_s' \mid < 2$
and $R_\eta$ was less than unity just because of
the difference of $p_T$ slope 
of semihard spectra
between midrapidity and forward/backward rapidity.
However, the initial parton density from the CGC
has no flat region.
Therefore, we should revisit the rapidity dependence of the
nuclear modification factor, because
partons travel almost straight line and
parton energy loss may directly probe the medium at each rapidity bin.

%%%%%%%%%%%%%%%%%%%%%%%%%%%%%%%%%%%%%%%%%%%%%%%%%%%%%%%%%%%%
% R_eta
%%%%%%%%%%%%%%%%%%%%%%%%%%%%%%%%%%%%%%%%%%%%%%%%%%%%%%%%%%%%
\begin{figure}[ht]
\includegraphics[width=3.5in]{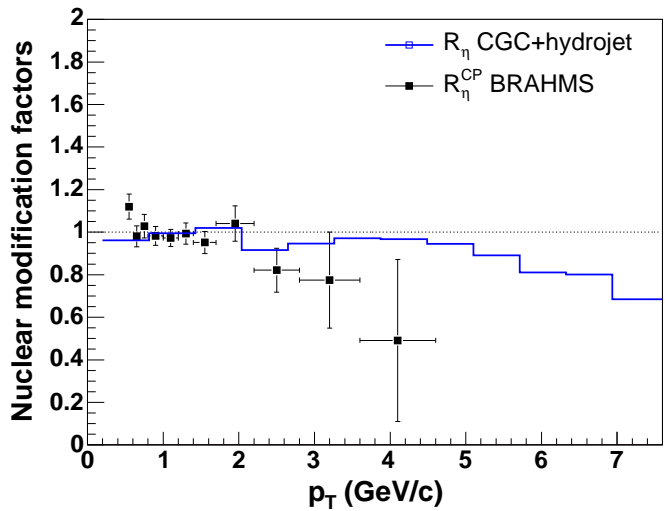}
\caption{
CGC + hydrojet result for $R_\eta$ is compared with
the BRAHMS data~\cite{BRAHMS:dA}
on $R_{\eta}^{\mathrm{CP}}
=R^{\mathrm{CP}}(\eta=2.2)/R^{\mathrm{CP}}(\eta=0)$.
Note that minijets are produced via
pQCD hard $2\rightarrow 2$ processes,
not $2 \rightarrow 1$.
}
\label{fig:reta}
\end{figure}
%%%%%%%%%%%%%%%%%%%%%%%%%%%%%%%%%%%%%%%%%%%%%%%%%%%%%%%%%%%%
The CGC+hydrojet result for charged hadrons
is compared with the BRAHMS data \cite{BRAHMS:dA}
in Fig.~\ref{fig:reta}.
We obtain $R_\eta \sim 1$ in the soft region $p_T<2$ GeV/$c$
which is dominated by the hydrodynamic component.
This is the same result as previous work even though
initial rapidity distribution is different.
The reason is the following.
Both initial conditions lead to reproduce the PHOBOS/BRAHMS data
on $dN_{\mathrm{ch}}/d\eta$
in which multiplicity at  $\eta \sim 2$ is almost the
same as at $\eta = 0$.
$p_T$ distributions at $\eta=0$ and 2.2 
are similar to each other below $p_T \sim 3$ GeV/$c$
in $pp$ collisions~\cite{HN3}.
We also find that the effect of radial flow
at $\eta_{\mathrm{s}} \sim 2$ is also
the same as at midrapidity.
Hence $R_{AA}$ in the low $p_T$ regions are insensitive to
pseudorapidity within this range.
One may worry about the effect of pseudorapidity
dependent Jacobian $J = p/E = \sqrt{1-m^2/(p_T^2 \cosh^2\eta+m^2)} $
on $p_T$ distributions.
However, this is already canceled in the calculation of $R_{AA}$
before obtaining its ratio $R_\eta$ in Eq.~(\ref{eq:reta}).

Situation is different in the high $p_T$ part.
our result starts to deviate from unity
at $p_T \sim 5$ GeV/$c$ since
$p_T$ slope from pQCD hard collisions
becomes much steeper for larger rapidity.
On the other hand,
experimental data deviates
prior to our result.
Note that the discrepancy between the present result
and our previous result \cite{HN3}
comes from the initial longitudinal shape
of the energy density shown in Fig.~\ref{fig:ini_dens} (a).
In the previous parametrization of the initial energy
density, dynamical evolution at midrapidity
was almost the same as at the {\em space-time rapidity} $\eta_{\mathrm{s}}=2$.
Thus $R_\eta<1$ came from purely
the difference of $p_T$ slope between midrapidity
and forward rapidity.
While, in the initial condition from the CGC,
initial energy density at midrapidity is
slightly larger than at $\eta_s = 2$.
Therefore, the ``slope" effect is compensated by
the dynamical effect.
This result
indicates that $p_T$ spectrum in forward
rapidity region is already suppressed
by the initial state effect.
If this is the case, instead of using conventional pQCD approach,
we may need to include $x$-evolution for the calculation
of $p_T$ spectrum up to $p_T=10$ GeV/$c$ that corresponds
to $x < 0.01$ at $y=2$~\cite{KLM03,Jalilian-Marian:2004xm}.
As demonstrated in Ref.~\cite{KLM04},
study of the back-to-back correlations at forward rapidity
will clarify the influence of parton saturation in the CGC.

\section{summary and outlook}

We developed a dynamical model for the description of
the time evolution of high dense matter created in heavy ion collisions
at high energies on the basis of the CGC in colliding nuclei,
hydrodynamic evolution for produced bulk matter, and
the parton energy loss in the medium.
In this approach, collisions of two CGC's are
calculated for the initial condition
of hydrodynamic simulations.
Thermalized dense partons produced from colliding two CGC's
are evolved hydrodynamically and also
used to evaluate energy loss of high-$p_T$ partons.

We have shown that
hydrodynamic simulations with the CGC initial conditions
describe centrality, rapidity,
 and energy dependences of charged hadron multiplicity
 very well. 
%
%centrality, rapidity, and energy dependences of charged hadron multiplicity
%can be very well described by the CGC initial condition in hydrodynamics.
%
The best agreement for the 
transverse momentum distribution was obtained
for pions
with the hydrodynamic component and the quenched hard jet component.
However, our results on kaons and protons indicated that
we need other mechanisms to understand
kaon and proton transverse momentum spectra from low to high-$p_T$.
The centrality dependence of nuclear modification factors
and the back-to-back correlation were found
to be consistent with the data up to the centrality of
$N_{\mathrm{part}} > 100$-150.
Finally we compared the nuclear modification factor at $\eta=2.2$
to BRAHMS data by using this new initial conditions.
We found that, when the QGP parton density
at the forward rapidity is smaller
than at mid-rapidity, suppression of hard jet from
pQCD calculation which is supplemented by DGLAP evolution
is not enough to account for the data.
It suggests that %pQCD calculation which correctly includes
the small-$x$ evolution will be necessary for the calculation
of particle spectra at forward rapidities at RHIC.

%
%  Outlook
%

Our main objective has been developing a consistent dynamical framework
for the time evolution of both soft matter and hard jets.
In order to meet this goal, many remains are to be done in the near future.
Each component should be improved to obtain a better description
of the whole of the stages of heavy ion collisions.

For initial conditions,
more sophisticated unintegrated gluon distribution should be used
as mentioned in Sec.~\ref{section:model}.
Numerical solutions of the classical Yang-Mills
equations in (2+1) dimension~\cite{KNV}
can be used for transverse profile in the boost invariant hydrodynamics.
Quantum evolution is important for the LHC
energy and forward/backward rapidity regions at RHIC.
(3+1)-D Yang-Mills solutions in which renormalization group evolution
of the color charges is implemented
are strongly demanded.

The ideal massless QGP EoS
 is employed in the deconfined phase
 in our present hydrodynamic model.
The EoS in hydrodynamics should be consistent with lattice QCD results.
For example, one can parametrize recent lattice results
by a resonance gas model for hadronic phase 
\cite{Karsch:2003zq}.
In connection with jet quenching phenomena,
it is argued that the small non-perturbative
screening mass
which is observed in lattice simulations \cite{Kaczmarek:1999mm}
leads to reduction of the amount
of jet quenching in the vicinity of phase transition
region \cite{Dumitru:2001xa}.
Therefore, one expect that
 jet quenching would strongly depend on the EoS~\cite{AD}.

One of the important goals in the jet tomography
is to constrain the evolving gluon density.
Therefore,  removing the free parameter $C$ in Eq.~(\ref{eq:GLV})
will be one of the most 
important future work within our approach
in order to see that
gluon density provided
by the hydrodynamic simulation is truly consistent with the
energy loss of partons.
In order to achieve this goal, we need to
extend our hydrodynamics and use more complicated formula
for the parton energy loss
at the same time.
For the more realistic description of chemical compositions
of quarks and gluons in the deconfinement phase,
we can solve hydrodynamic equations
together with rate equations for number
densities of quarks and
gluons \cite{Biro:1993qt}.
This also affects jet quenching through the different
 cross section among $gg$, $gq$, and $qq$ processes.
The kinematic effect of radiated gluons,
fluctuations, or higher order terms
in the opacity expansion in GLV formula
 should be taken into account
in the energy loss of hard partons.

\begin{acknowledgements}
The authors are grateful to A.~Dumitru,
U.~Heinz, J.~Jalilian-Marian,  D.~Kharzeev, L.~McLerran,
D.~Teaney, and R.~Venugopalan
for useful comments.
The work of T.H. was  supported by RIKEN.
Research of Y.N. was supported by the DOE under Contract No.~DE-FG03-93ER40792.
\end{acknowledgements}

\end{document}